\documentclass[12pt]{article}
\pdfoutput=1
\usepackage{amsmath,amssymb,amsfonts}
\usepackage[pdftex]{graphicx}
\usepackage{epsfig}
\usepackage{verbatim}
\usepackage[margin=1cm]{caption}

\makeatletter \@addtoreset{equation}{section} \makeatother
\makeatletter \@addtoreset{figure}{section} \makeatother

\def\CA{{\cal A}}\def\CB{{\cal B}}
\def\CE{{\cal E}}\def\CF{{\cal F}}

\def\CK{{\cal K}}\def\CL{{\cal L}}\def\CM{{\cal M}}
\def\CN{{\cal N}}\def\CO{{\cal O}}\def\CP{{\cal P}}
\def\CQ{{\cal Q}}\def\CR{{\cal R}}\def\CS{{\cal S}}
\def\CT{{\cal T}}\def\CV{{\cal V}}
\def\CW{{\cal W}}\def\CX{{\cal X}}
\def\CC{{\cal C}}

\def\a{\alpha}\def\b{\beta}\def\g{\gamma}
\def\d{\delta}\def\e{\epsilon}
\def\z{\zeta}\def\th{\theta}
\def\k{\kappa}\def\l{\lambda}
\def\m{\mu}
\def\r{\rho}\def\s{\sigma}
\def\t{\tau}
\def\w{\omega}\def\G{\Gamma}
\def\D{\Delta}\def\L{\Lambda}
\def\S{\Sigma}

\newcommand{\be}{\begin{equation}}
\newcommand{\ee}{\end{equation}}
\newcommand{\bea}{\begin{eqnarray}}
\newcommand{\eea}{\end{eqnarray}}
\newcommand{\ba}{\begin{array}}

\def\ba{{\bar a}}
\def\be{{\bar e}}

\begin{document}
\begin{titlepage}
\vfill
\begin{flushright}
{\tt\normalsize KIAS-P13059}\\
{\tt \normalsize EFI-13-28}\\

\end{flushright}
\vfill
\begin{center}
{\Large\bf Exact Partition Functions on $\mathbb{RP}^2$  and Orientifolds}

\vskip 1cm

Heeyeon Kim\footnote{\tt hykim@phya.snu.ac.kr}$^\dagger$,
Sungjay Lee\footnote{\tt sjlee79@uchicago.edu}$^\diamond$,
and Piljin Yi\footnote{\tt piljin@kias.re.kr}$^\ddagger$

\vskip 5mm
$^\dagger${\it Department of Physics and Astronomy, Seoul
National University, \\Seoul 151-147, Korea}
\vskip 3mm $^\diamond${\it Enrico Fermi Institute and Department of Physics
University of Chicago \\
5620 Ellis Av., Chicago Illinois 60637, USA}
\vskip 3mm $^\ddagger${\it School of Physics, Korea Institute
for Advanced Study, Seoul 130-722, Korea}

\end{center}
\vfill

\begin{abstract}
\noindent
We consider gauged linear sigma models (GLSM) on $\mathbb{RP}^2$,
obtained from a parity projection of $S^2$. The theories admit
squashing deformation, much like GLSM on $S^2$, which allows us to
interpret the partition function as the overlap amplitude
between the vacuum state and crosscap states. From these,
we extract the central charge of Orientifold planes, and observe that
the Gamma class makes a prominent appearance as in the
recent D-brane counterpart. We also repeat the computation
for the mirror Landau-Ginzburg theory, which naturally brings
out the $\theta$-dependence as a relative sign between
two holonomy sectors on $\mathbb{RP}^2$. We also
show how our results are consistent with known topological
properties  of D-brane and Orientifold plane world-volumes,
and discuss what part of the wrapped D-brane/Orientifold
central charge should be attributed to the quantum volumes.

\end{abstract}

\vfill
\end{titlepage}

\parskip 0.1 cm
\setcounter{tocdepth}{2}
\tableofcontents\newpage
\renewcommand{\thefootnote}{\#\arabic{footnote}}
\setcounter{footnote}{0}

\parskip 0.2 cm

\section{Introduction and Summary}

The two-dimensional (2,2) gauged linear sigma model (GLSM) is
a very useful tool for studying conformal field theories of
Calabi-Yau manifolds \cite{Witten:1993yc}. It allows us to
understand how the large volume limit is smoothly connected
to Landau-Ginzburg descriptions, and provides an intuitive and
straightforward proof of the mirror symmetry.

The most notable development involving GLSM in recent years,
by far, is the formulation of GLSM on $S^2$ and on squashed $S^2$,
and the computation of the partition function thereof. As
conjectured initially in Ref.~\cite{Jockers:2012dk} and
argued from the squashed $S^2$ versions in Ref.~\cite{Gomis:2012wy},
this leads to a new way to compute exact geometry
of K\"ahler moduli space, without referring to the
mirror symmetry dual. The conjectured relationship to the A-model
$tt^*$-amplitude  with K\"ahler parameters, or the complexified
FI parameters $\tau$, is
\begin{equation}
e^{-\CK(\tau,\bar\tau)}= {}_{R}\langle 0|\bar0\rangle_{R}=Z_{S^2}(\tau,\bar\tau)\ ,
\end{equation}
where ${}_{R}\langle 0|$ is a canonical ground state of
the Ramond sector \cite{Cecotti:1991me},
$Z_{S^2}$ is a two-sphere partition function
of (2,2) GLSM which was calculated exactly using
supersymmetric localization at \cite{Doroud:2012xw,Benini:2012ui}.
This provides a direct method of computing Gromov-Witten
invariants, i.e., the world-sheet instanton contribution
to the above quantity, in a manner that obviates the mirror B-model.

This in turn leads to another natural question of how
boundary state amplitudes are computed in this new
approach. Refs.~\cite{Hori:2013ika,Sugishita:2013jca,Honda:2013uca} extended the
above to a hemisphere partition function. Interestingly,
the supersymmetry that survives the squashing of $S^2$ is
such that it is naturally A-twisted (anti-A-twisted) at
the poles but at the same time B-twisted at the equator. Thus, the
boundary states one can attach to the hemisphere are
holomorphic cycles wrapped by D-branes. Along the same
logic as above, the hemisphere partition function then
computes the overlap amplitude between the canonical vacuum
and the boundary states in question,
\begin{equation}\label{centralcharge}
\Pi_{B}^0(\tau)={}_R\langle0|B\rangle_R=Z_{D^2}(B,\tau)\ ,
\end{equation}
which is nothing but the central charge of the D-brane.

One of more interesting results from this can be seen from
the large volume limit. Explicit results for simple hypersurface
examples, say that the central charge in the large
volume limit is
\begin{equation}
\int_\CX e^{-B-iJ}\wedge ch(\CF)\wedge
\hat\Gamma_c(\CT)\ ,
\end{equation}
where $\hat\Gamma_c({\CT})$ is a multiplicative characteristic
class defined by
\begin{equation}
\hat\Gamma_c({\CT}) = \prod_{i=1}^{d}\Gamma\left(1+\frac{x_i}{2\pi i}\right)\ ,
\end{equation}
and $\CT$ is the (holomorphic) tangent bundle of the Calabi-Yau.
Most notably, this corrects the conventional form of the
central charge as
\begin{equation}
\sqrt{\CA(\CT)} \quad \rightarrow\quad \hat\Gamma_c({\CT})\ .
\end{equation}
This appearance of $\hat\Gamma_c$ class has been foretold
from various explicit computations via mirror symmetry \cite{Libgober,Iritani,Katzarkov:2008hs,Halverson:2013qca}.

In this note, we initiate extending these works to the presence of
Orientifold planes. The simplest quantity one can compute is the
vacuum-to-crosscap amplitude,
\begin{equation}\label{O1}
{}_R\langle0|{\CC}\rangle_R\ .
\end{equation}
Pictorially, this  is computed by a cigar-like geometry with
the identity operator at the tip and a crosscap at the other end.
There are  two possible choices for the crosscap, say,
A-type and B-type. The former corresponds to Orientifold planes
wrapping Lagrange subcycles.
In this note, we are led to consider B-type parity for GLSM, for
much the same reason as in Ref. \cite{Brunner:2003zm}, which
corresponds to Orientifold planes wrapping the holomorphic cycles.
Topologically the world-sheet is that of $\mathbb{RP}^2$, and
the same squashing deformation as in Ref.~\cite{Gomis:2012wy}
is allowed, the partition function of GLSM on $S^2/Z_2 =\mathbb{RP}^2$
is expected to compute the vacuum-to-crosscap amplitude,
\begin{equation}\label{centralcharge}
{}_R\langle0|\CC\rangle_R=Z_{\mathbb{RP}^2}(O,\tau)\ .
\end{equation}
In the convention of Brunner-Hori \cite{Brunner:2003zm},
the relevant parity action for our purpose here is of type B, which
leads to, generally, holomorphically embedded Orientifold planes.
Computation of the partition function follows easily from the
$S^2$ partition function computation, and the result is expressed
in terms of a product of the Gamma functions. See section 3 for the complete expressions.

For Orientifold plane that wraps the Calabi-Yau $\CX$ entirely,
we also take the large volume limit of the central charge.
Conventionally, Orientifold planes, $O^\pm$, have $\CL^{1/2}$ class
as the counterpart of D-branes' $\CA^{1/2}$ class. Here we find
that one must also replace
\begin{equation}\label{11}
\sqrt{\CL(\CT/4)}\quad\rightarrow\quad \frac{{\cal A}({\cal T}/2)}{\hat\Gamma_c(-\CT)}\ .
\end{equation}
The parity action
on $S^2$ can be augmented by additional $Z_2$ action on the chiral
fields, which induces various combinations of $O_{2(d-s)}$ planes,
say wrapping a submanifold $\CM$. For these cases, we must also replace
\begin{equation}\label{sub}
\sqrt{\frac{\CL(\CT/4)}{\CL(\CN/4)}}\quad\rightarrow\quad
\frac{{\cal A}({\cal T}/2)}{\hat\Gamma_c(-\CT)}\wedge
\frac{\hat\Gamma_c(\CN)}{{\cal A}({\cal N}/2)}
\ ,
\end{equation}
with the normal bundle $\CN$ and the tangent bundle $\CT$ of
holomorphically embedded  $\CM$ in the Calabi-Yau $\CX$.
For more complete expression for the large volume limit,
see section 5.

The results found here should be consistent with the
hemisphere computation of the D-brane central charges.
Among those issues discussed are anomaly inflow and a
twist that is known to be present when the world-volume
wraps a Spin$^c$ (rather than Spin, i.e.) submanifold.
Also, one  outfall
from having both D-brane and Orientifold plane central
charges available is the interpretation of exactly what
the Gamma class corrects. The central charge does not by itself
tells us whether the correction goes to the RR-charge or
the vacuum expectation values of spacetime scalars, or
equivalently the quantum volumes. Our conclusion
is that the correction should be attributed entirely
to the $\alpha'$ correction of volumes.

This note is organized as follows. Section 2 outlines GLSM on
$\mathbb{RP}^2$ as a type B-parity projection of that on $S^2$, and
briefly discusses squashing deformation of $\mathbb{RP}^2$
to motivate the interpretation of the $\mathbb{RP}^2$ partition
functions as vacuum-to-crosscap amplitudes.
Section 3 computes the partition function exactly: We start with
identification of two saddle points, of even and odd holonomy
respectively, and compute the relevant 1-loop determinants of
chiral and vector multiplets. The parity action can be accompanied
by $Z_2$ flavor rotations, which correspond to Orientifold planes
of even co-dimensions. In section 4, we turn to the mirror
Landau-Ginzburg description and recover the results of section 3.
Here we also learn how the two possible values of $\theta$ angle, i.e.
$\theta=0,\pi$, affect the partition functions and sometimes
distinguish the (relative) type of Orientifolds from different
holonomy sectors.

Section 5 specializes the result to the case of Calabi-Yau hypersurface
$\CX$ in $\mathbb{CP}^{N-1}$, and various Orientifolds
thereof, and extracts the perturbative contribution. This gives
the large volume expression of the central charge, where
the $\hat\Gamma_{c}$ class makes appearance as in \eqref{11} and \eqref{sub}.
Section 6 will consider subtleties and make some
consistency checks, from the simple tadpole condition to anomaly inflow.
The latter in particular suggests that topological content of D-branes
and Orientifold planes remain unchanged despite the changes in the
central charges. We point out that, in all central charge expressions
from the hemisphere and $\mathbb{RP}^2$ partition functions,
the multiplicative  shift due to the appearance of $ \hat\Gamma_c$
class must be  understood as quantum shift of $e^{-iJ}$,
such that the RR-charges and the Chern-Simon couplings remain unchanged.

In the Appendices, we outline some technical aspects of the computation
but also address a well-known subtlety when $\CM$ is a proper submanifold
of Spin$^c$ structure. Invoking tachyon condensation, we motivate natural
R-charge and gauge charge assignment for the boundary Hilbert space, how
an extra factor $e^{-c_1(\CN)/2}$ emerges for D-branes when $\CM$ is
a proper and Spin$^c$ submanifold.

\section{GLSM on $\mathbb{RP}^2$ and Squashing}

In this section, we start with a brief review on general aspects of
parity symmetries in 2d (2,2) theory on $\mathbb{R}^{1+1}$,
which were thoroughly studied in Ref.~\cite{Brunner:2003zm}.
To begin with, the parity action on the 2-dimensional superspace
$(x^\pm=x^0\pm x^1, \theta^{\pm},\bar\theta^{\pm})$ is
$x^1 \rightarrow -x^1$, accompanied by the proper action
in the fermionic coordinates. Depending on the latter
there are two distinct possibilities,
\begin{eqnarray}\nonumber
\Omega_A &:& (x^\pm,\theta^\pm, \bar\theta^\pm) \rightarrow
(x^\mp, -\bar\theta^\mp, -\theta^\mp)\ , \\
\Omega_B &:& (x^\pm,\theta^\pm, \bar\theta^\pm) \rightarrow
(x^\mp, \theta^\mp, \bar\theta^\mp)\ ,
\end{eqnarray}
which we will call A and B-parity respectively. Under this action,
the four supercharges transform as
\begin{eqnarray}\nonumber
A &:& Q_\pm \rightarrow  \bar Q_\mp, ~~\bar Q_\pm \rightarrow Q_\mp,\\
\nonumber
&& D_\pm \rightarrow \bar D_\mp, ~~\bar D_\pm \rightarrow D_\mp,\\
\nonumber
B &:& Q_\pm \rightarrow  Q_\mp, ~~\bar Q_\pm \rightarrow \bar Q_\mp,\\
&& D_\pm \rightarrow D_\mp, ~~\bar D_\pm \rightarrow \bar D_\mp\ .
\end{eqnarray}
Hence, under the A-parity action, half of the supersymmetry
is broken, leaving $Q_A\equiv Q_+ + \bar Q_-$ and $Q_A^\dagger$ invariant.
Under B-parity, and $Q_B\equiv \bar Q_++\bar Q_-$ and $Q_B^\dagger$
survive.

Furthermore, the simplest transformation rule for a chiral field
$(\phi,\psi,F)$ is
\begin{eqnarray}\nonumber
A&:&\phi(x) \rightarrow \bar\phi(x')\ ,\\\nonumber
&&\psi_\pm(x) \rightarrow \bar\psi_\mp(x')\ ,\\
&&F(x)\rightarrow \bar F(x')\ ,\\\nonumber
B&:& \phi(x) \rightarrow \phi(x')\ ,\\\nonumber
&& \psi_\pm(x) \rightarrow \psi_\mp(x')\ ,\\\label{projection}
&&F(x) \rightarrow -F(x')\ ,
\end{eqnarray}
and one can check that these leave the kinetic lagrangian of
the chiral multiplet invariant. For a twisted chiral multiplet,
transformation rules under A and B-parities are exchanged.

For each parity projection, we can associate a crosscap
state denoted by $|\CC_{A,B}\rangle$. Then we can think of the
overlap between this state and a (twisted) chiral ring element,
such as
\begin{equation}
\langle a|\CC_B\rangle\ .
\end{equation}
We naturally expect that this quantity calculates
the Orientifold analogue of the D-brane central charge.
Among these overlaps, there are distinguished element
$\langle 0|\CC_B\rangle$ that no chiral field is inserted
at the tip of the hemisphere. The path integral can be done
by doubling of the hemisphere by gluing its mirror
image. Topology of the world-sheet is that of a two
sphere with antipodal points identified, i.e., $\mathbb{RP}^2$.

\subsection{GLSM on $\mathbb{RP}^2$}

The supersymmetric Lagrangian we are considering is the same as that used in
\cite{Doroud:2012xw,Benini:2012ui};
\begin{eqnarray}
\CL = \CL_{vector}+\CL_{chiral}+\CL_{W}+\CL_{FI}\ ,
\end{eqnarray}
where the kinetic terms for the vector and the charged chiral multiplets are,
respectively,
\begin{align}
\CL_{vector} = & \ \frac{1}{2g^2}
\mathrm{Tr}\left[\left(F_{12}+\frac{\sigma_1}{r}\right)^2
+(D_\m \sigma_1)^2
+(D_\m \sigma_2)^2
-[\sigma_1,\sigma_2]^2
+D^2\right.
\nonumber \\
&+i\bar\lambda\gamma^\m D_\m \lambda
+i\bar\lambda[\sigma_1,\lambda] +i\bar\lambda\g^3[\sigma_2,\lambda]\Bigg]\ ,
\label{vector}\\
\CL_{chiral} = & \ \bar\phi\left(-D^\m D_\m+\sigma_1^2+\sigma_2^2+iD
+i\frac{q-1}{r}\sigma_2+\frac{q(2-q)}{4r^2}\right)\phi +\bar F F
\nonumber \\ \label{chiral}
&-i\bar\psi\left(\gamma^\m D_\m-\sigma_1-i\gamma^3\sigma_2
+\frac{q}{2r}\gamma^3\right)\psi+i\bar\psi\lambda\phi-i\bar\phi\bar\lambda\psi\ ,
\end{align}
and the potential terms take the following form,
\begin{eqnarray}
\CL_\CW = \sum_i\frac{\partial \CW}{\partial\phi^i}F^i
-\frac12\sum_{i,j}\frac{\partial^2\CW}{\partial\phi^i\partial\phi^j}\psi^i\psi^j
+ \text{c.c.}\ .
\end{eqnarray}
Finally the Fayet-Illiopoulos (FI) coupling and the two-dimensional topological term
are
\begin{eqnarray}
\CL_{FI} = -\frac{\t}{2} \text{Tr} \Big[D - \frac{\s_2}{r} +i F_{12}\Big] + \frac{\bar{\t}}{2}
\text{Tr}\Big[D- \frac{\sigma_2}{r} -iF_{12}\Big]\ ,
\end{eqnarray}
where $\t=i\xi+\frac{\theta}{2\pi}$, ($\xi\in\mathbb{R}$, $\theta\in[0,2\pi]$).
Note that the superpotential $\CW(\phi)$
should carry $R$-charge two to preserve the supersymmetry on $\mathbb{RP}^2$.

The Lagrangian is invariant under the supersymmetry transformation rules,
\begin{align}
  \d \l = & (i V_1 \g^1 + iV_2 \g^2 + iV_3 \g^3    - {\rm D}) \e \ ,
  \nonumber \\
  \d \bar \l = & ( i \bar V_1 \g^1 + i \bar V_2 \g^2 + i \bar V_3 \g^3
  + {\rm D}) \bar \e    \ , \nonumber\\
  \d A_i = & - \frac i2 \Big( \bar \e \g_i \l - \bar \l \g_i \e \Big) \ ,  \nonumber
  \\
  \d \s_1 = &   \frac12 \Big( \bar \e \l - \bar \l \e \Big)\ ,
  \nonumber \\
  \d \s_2 = & - \frac i2 \Big( \bar \e \g_3 \l - \bar \l \g_3 \e \Big)\ ,
  \nonumber \\
  \d D = & - \frac i2 \bar \e \g^\m D_\m  \l - \frac i2 \big[ \s_1, \bar \e \l \big]
  - \frac12 \big[ \s_2 , \bar \e \g^3 \l \big]\ ,
  \nonumber \\ &
  + \frac i2 \e \g^\m D_\m  \bar \l
  - \frac i2 \big[ \s_1 , \bar \l \e \big] - \frac12 \big[ \s_2 , \bar \l \g^3 \e \big]\ ,
  \label{SUSYvec}
\end{align}
with
\begin{align}
  \vec V \equiv & \Big( + D_1 \s_1 + D_2 \s_2 ,\  + D_2 \s_1 - D_1 \s_2 ,
  \ F_{1 2} + i [\s_1,\s_2] + \frac{1}{r}\s_1 \Big)\ ,
  \nonumber \\
  \vec{\bar V} \equiv & \Big( - D_1 \s_1 + D_2 \s_2 ,\ - D_2 \s_1 - D_1 \s_2 ,
  \ F_{1 2} - i [\s_1,\s_2] + \frac{1}{r}\s_1 \Big)\ ,
\end{align}
and
\begin{align}
  \d \phi = & \bar \e \psi \ \ ,   \nonumber \\
  \d \bar \phi = & \e \bar \psi \ , \nonumber \\
  \d \psi = & i \g^\m \e D_\m \phi + i\e \s_1 \phi + \g^3 \e \s_2 \phi + i \frac{q}{2r}
  \g_3 \e \phi + \bar \e F\ \ ,  \nonumber \\
  \d \bar \psi = & i  \g^\m\bar \e  D_\m \bar \phi + i \bar \e \bar \phi \s_1
  - \g^3 \bar \e  \bar \phi \s_2 - i \frac{q}{2r} \g_3 \bar \e  \bar \phi + \e \bar F\
  \nonumber \ , \\
  \d F  = & \e \Big( i \g^i D_i \psi  - i \s_1 \psi + \g^3 \s_2 \psi - i \l \phi \Big)
  - i \frac{q}{2} \psi \g^i D_i \e \   \ , \nonumber \\
  \d \bar F  = & \bar \e \Big( i \g^i D_i \bar\psi - i\bar\psi \s_1 - \g^3 \bar \psi \s_2
  + i \bar \phi \l \Big) - i \frac{q}{2} \bar \psi \g^i D_i \bar \e \ .
  \label{SUSYmatter}
\end{align}
Here the spinors $\e$ and $\bar \e$ are given by\footnote{See Appendix A for our
gauge choice.}
\begin{eqnarray}\label{killing}
\epsilon =
e^{i\varphi/2}\left(\begin{array}{c}\cos\theta/2\\\sin\theta/2\end{array}\right)\ ,
\qquad
\bar\epsilon =e^{-i\varphi/2}\left(\begin{array}{c}\sin\theta/2\\
\cos\theta/2\end{array}\right)\ ,
\end{eqnarray}
satisfying the Killing spinor equations
\begin{align}
  \nabla_\m\e = \frac{1}{2r} \g_\m \g^3 \e \ ,
  \qquad
  \nabla_\m \bar \e = - \frac{1}{2r} \g_\m \g^3 \bar \e \ .
\end{align}
Note that the surviving  supersymmetry (\ref{killing}) becomes A-type and B-type supersymmetry
at the pole ($\th=0$) and the equator ($\th=\pi/2$), respectively.

In order to define the theory on $\mathbb{RP}^2$, we further impose
a suitable parity projection condition on the dynamical fields so that
the Lagrangian is invariant under the parity.
Particularly, one has to consider the type B-parity in the following
discussion. This is because the Killing spinors (\ref{killing})
transform as
\begin{align}
  \epsilon_\pm \rightarrow i\epsilon_\mp\ ,
  \qquad
  \bar\epsilon_\pm \rightarrow -i\bar\epsilon_\mp\ ,
  \label{bparitykilling}
\end{align}
under the parity action
$(\theta,\varphi)\rightarrow (\pi-\theta,\varphi+\pi)$.
It implies that the B-type Orientifold plane can be naturally
placed at the equator $\th=\pi/2$.

We remark here that, as in the case of the $S^2$, the Lagrangian except
$\CL_{FI}$ can be made $Q$-exact with the supersymmetry chosen by
(\ref{killing}). For instance,
\begin{align}
  \CL_{vector} =  \frac{1}{g^2} \d_\e \d_{\bar \e} \text{Tr}\Big[
  \frac12 \bar \l \g^3\l - 2i D \s_2 + \frac ir \s_2^2 \Big]\ ,
\end{align}
and
\begin{align}
  \CL_{chiral}=  - \d_\e \d_{\bar \e} \Big[ \bar \psi \g^3 \psi
  - 2 \bar \phi (\s_2 + i\frac{q}{2r} ) \phi + \frac ir \bar \phi \phi \Big]\ .
\end{align}
Consequently, the partition function on $\mathbb{RP}^2$
contains only the A-model data.

\subsection{Squashed $\mathbb{RP}^2$ and Crosscap Amplitudes}

We propose that the partition function
of $\CN=(2,2)$ GLSM on $\mathbb{RP}^2$ computes
the overlap between the supersymmetric ground state and the type B-crosscap
state in the Ramond sector
\begin{align}
  Z_{\mathbb{RP}^2} = ~_{R}\langle 0 | \CC_B \rangle_{R}\ ,
\end{align}
which is the central charge of the Orientifold plane.
To understand the above proposal, it is useful to consider a squashed
$\mathbb{RP}^2$, denoted by $\mathbb{RP}^2_b$,
where the Hilbert space interpretation of the results in section 3 becomes clear.

The squashed $\mathbb{RP}^2$ can be described by
\begin{align}
  \frac{x_1^2+x_2^2}{l^2} + \frac{x_3^2}{\tilde l^2} = 1
\end{align}
with $Z_2$ identification below
\begin{align}
  Z_2 \ : \ \left( x_1,x_2,x_3\right) \ \to \ \left( -x_1, -x_2,-x_3 \right)\ .
\end{align}
The metric on this space is
\begin{align}
  ds^2 = f^2(\th) d\th^2 + l^2 \sin^2\th d\varphi^2 \ ,
\end{align}
where $f^2(\th) = \tilde l^2 \sin^2\th + l^2 \cos^2\th$. The world-sheet parity
$Z_2$ acts on the polar coordinates as follows
\begin{align}
  Z_2 \ : \ \left( \th,\varphi\right) \ \to \left( \pi - \th, \pi+\varphi \right) \ .
\end{align}
An Orientifold plane is placed at the equator $\th=\pi/2$.
By turning on a suitable
background gauge field coupled to the $U(1)_V$ current,
\begin{align}
  V = \frac12 \left( 1- \frac{l}{f(\th)} \right) d\varphi\ ,
  \label{Vr}
\end{align}
valid in the region $0<\th<\pi$, one can show the Killing spinors (\ref{killing})
on the squashed $\mathbb{RP}^2$ satisfying the generalized Killing spinor equations
\begin{align}
  D_m \e = \frac{1}{2f} \g_m \g^3 \e \ ,
  \qquad
  D_m \bar \e = -\frac{1}{2f} \g_m \g^3 \bar \e \ ,
  \label{modifiedKS}
\end{align}
where the covariant derivative denotes $D_m=\partial_m - i V_m$. Here we normalize
the $R$-charge so that the Killing spinor $\e$ ($\bar \e$) carries $+1$ ($-1$) $R$-charge.

As in Ref.~\cite{Gomis:2012wy},
one can show that the partition function is invariant no matter how much we squash
the space $\mathbb{RP}^2$, i.e., it is independent of the squashing parameter $b=l/\tilde l$.
Appendix \ref{squashing} shows detailed computations for this.
In the limit $b\to 0 $, we have an infinitely stretched cigar-like
geometry where the type B-crosscap state $|\CC_B\rangle$ is prepared at $\th=\pi/2$.
Near $\th\simeq {\pi}/{2}$,
all the fields can be made periodic along the circle $S^1$
due to the background gauge field $V\simeq \frac12d\varphi$, which implies that
the theory is in the Ramond sector near $\th \simeq \pi/2$.
Moreover, as mentioned earlier, the partition function on the squashed $\mathbb{RP}^2$ contains
only the A-model data.

Combining all these facts,
we can identify the partition function on $\mathbb{RP}^2_b$
as the overlap in the Ramond sector between A-model ground state corresponding to the identity operator at the tip
and the B-type crosscap state defined by an appropriate projection condition we discuss soon,
\begin{align}
  Z_{\mathbb{RP}^2} = Z_{\mathbb{RP}^2_b} \overset{b\to0}{=} ~_{R}\langle 0 | \CC_B \rangle_{R}\ .
\end{align}

\section{Exact $\mathbb{RP}^2$ Partition Function}

In this section, we compute the partition function of GLSM on $\mathbb{RP}^2$ exactly,
via the  localization technique. The analysis is parallel to the computation
of the two-sphere partition function \cite{Doroud:2012xw,Benini:2012ui}.

As we will be working with the Coulomb phase saddle points,
the gauge group is effectively reduced to the Cartan subgroup $U(1)^{r_G}$, whose
scalar partners will be collectively denoted by $\sigma$.
The relevant gauge charges are expressed via weights and roots.
For chiral multiplets in the $G$-representation $\mathbf{R}$,
 these $U(1)^{r_G}$ gauge charges
will be denoted collectively as $w$, so the 1-loop determinant
of a chiral multiplet with weight $w$ is a function of
$w \cdot\sigma$. When the gauge group is Abelian as in sections 4, 5, and 6,
we also use the notation $Q$ for the gauge charges, so $w\cdot\sigma$
is written as $Q\cdot \sigma$. Similarly, contribution from each massive
``off-diagonal" vector multiplet is determined entirely
by its charge under the unbroken $U(1)^{r_G}$;
the determinant is then written in terms of $\alpha\cdot\sigma$.
In the end, we take a product over all the weights, $w$,
and all the roots, $\alpha$.

\subsection{Saddle Points}

To apply the localization technique, we choose the kinetic terms
$\CL_{vector}$ and $\CL_{chiral}$ as the $Q$-exact deformation and scale them
up to infinitely.
The path-integral then localizes at the supersymmetric saddle points
satisfying the equations
\begin{equation}
F_{12}=-\frac{\sigma_1}{r}=\frac{B}{2r^2},~~
D_\mu\sigma_1=D_\mu\sigma_2=[\sigma_1,\sigma_2]=0,~~
D+\frac{\sigma_2}{r}=0\ ,
\label{saddle1}
\end{equation}
with all the other fields vanishing. Among these saddle configurations,
the only one invariant under the B-type Orientifold projection is
\begin{equation}
F_{12}=0,~~\sigma_1=0,~~D_\mu\sigma_2=0,~~D+\frac{\sigma_2}{r}=0\ .
\label{saddle2}
\end{equation}

However, since $\mathbb{RP}^2$ has a non-contractible loop $C$ which
connects two antipodal points in the equator, $F_{12}=0$ is solved
by a flat connection with a discrete $Z_2$ holonomy
\begin{equation}
\CP \exp\left[i\int_C A \right] \in Z_2 \ .
\end{equation}
Hence there are two kinds of saddle points, which we call even and odd
holonomy. Near the odd holonomy, fields effectively satisfy twisted
boundary condition that picks up additional sign along the loop.

Finally, using $U(N)$ gauge transformation, we can make $A_\mu$
holonomy and constant mode of $\sigma_2$
both diagonal, as the two must commute with each other. Then
the saddle point configurations all reduce to
\begin{equation}
\sigma_2 = \sigma, ~~D=-\frac{\sigma}{r}\ ,
\end{equation}
where $\sigma$ is arbitrary constant element in the Cartan subalgebra.
The classical action at the saddle points is,
\begin{equation}
Z_{classical}=e^{-i2\pi r\xi \sigma}  .
\end{equation}

\subsection{Chiral Multiplets}

In this section, we calculate  one-loop determinants of  chiral
multiplets, say, in the representation $\mathbf{R}$ of the gauge group $G$.
To compute the one-loop determinant, we truncate the regulator action
up to quadratic order in small fluctuation, around each saddle point
\begin{align}
  S_{chiral} = S_{chiral}^b + S_{chiral}^f \nonumber
\end{align}
with
\begin{align}
  S_{chiral}^b = \int d^2x \sqrt{g} \ \bar\phi \Big[ - D_\m^2 + \s^2
  +i\frac{q-1}{r}\sigma + \frac{q(2-q)}{4r^2} \Big] \phi \ ,
\end{align}
and
\begin{align}
  S_{chiral}^f = \int d^2x \sqrt{g} \ \bar\psi \g^3 \Big[ - i \g^3\g^\m D_\m
  -  \left( \s + i \frac{q}{2r}\right) \Big] \psi \ .
\end{align}
We refer readers to Appendix \ref{harmonics} for properties
of the relevant spherical harmonics.

\paragraph{Even Holonomy}

First, we will calculate the contribution near the first saddle point,
where the holonomy is trivial.
For this, we impose the  B-type Orientifold
projection\footnote{This choice of projection condition is consistent
with the supersymmetry (\ref{SUSYmatter}) and (\ref{killing}).
},
\begin{align}
  \phi (\pi-\th,\pi+\varphi) = & + \phi(\th, \varphi)\ ,
  \nonumber \\
  \psi_\pm ( \pi -\th, \pi+\varphi) = & - i \psi_\mp (\th,\varphi) \ ,
  \nonumber \\
  \bar \psi_\pm (\pi-\th,\pi+\varphi) = & + i \bar \psi_\mp (\th,\varphi)\ ,
  \nonumber \\
  F (\pi -\th, \pi+\varphi) = & + F (\th,\varphi) \ .
  \label{chiralproj}
\end{align}
For simplicity, let us first consider a single chiral multiplet
of charge $+1$ under a $U(1)$ gauge group. Thanks to the property,
with our gauge choice,
\begin{align}
  Y_{\mathbf{q},jm}(\pi-\th,\pi+\varphi) = & (-1)^{j} e^{-i\pi|\mathbf{q}|}
  Y_{-\mathbf{q},jm}(\th,\varphi)\ ,
\end{align}
we can write scalar fluctuations that survive under the projection (\ref{chiralproj}) as
\begin{eqnarray}
\phi(\theta,\varphi)=\sum_{\substack{j=2k\\ k\geq 0}} \sum_{m=-j}^{j} \phi_{jm} Y_{jm}\ .
\end{eqnarray}
The bosonic part of the quadratic action then becomes
\begin{align}
  S_{chiral}^b = \frac12 \sum_{\substack{j=2k\\ k\geq 0}}\sum_{m=-j}^{j}
  \bar \phi_{jm} \Big[ \left( j + \frac q2 - i r \s \right)
  \left( j+1-\frac q2 + i r \s \right) \Big] \phi_{jm}\ ,
\end{align}
which leads to
\begin{equation}
\mathrm{Det}_\phi =\prod_{k\geq 0}
\left(2k+\frac{q}{2}-ir\sigma\right)^{4k+1}
\left(2k+1-\frac{q}{2}+ir\sigma\right)^{4k+1}\ .
\end{equation}
Next, the mode expansion of the fermion fluctuation invariant
under the projection (\ref{chiralproj}) takes the form
\begin{align}
  \psi = & \sum_{\substack{j=2k+1/2 \\ k\geq0}} \sum_{m=-j}^j \psi^+_{jm} \Psi^+_{jm}
  + \sum_{\substack{j=2k+3/2 \\ k\geq0}} \sum_{m=-j}^j \psi^-_{jm} \Psi^-_{jm}\ ,
  \nonumber \\
  \bar \psi = & \sum_{\substack{j=2k+1/2 \\ k\geq0}}
  \sum_{m=-j}^j \bar \psi^+_{jm} \bar\Psi^+_{jm}
  + \sum_{\substack{j=2k+3/2 \\ k\geq0}} \sum_{m=-j}^j \bar \psi^-_{jm} \bar\Psi^-_{jm}\ ,
\end{align}
where the spinor harmonics $\Psi^\pm_{j,m}$ are
\begin{align}
  \Psi^\pm_{jm} = \begin{pmatrix}   Y_{-\frac 12,jm}
  \\ \pm Y_{\frac 12, jm}  \end{pmatrix}\ ,
  \qquad
  \bar\Psi^\pm_{jm} = \begin{pmatrix} Y^*_{\frac 12,jm}
  \\ \pm Y^*_{-\frac 12, jm}  \end{pmatrix} \ .
\end{align}
In terms of the mode variables, the fermionic part of the quadratic action can be
expressed as
\begin{align}
  S_{chiral}^f = & + i \sum_{\substack{j=2k+1/2 \\ k\geq0}} \sum_{m=-j}^j
  \bar\psi^+_{jm} \Big[ j+ \frac12 - \frac q2 + i r \s \Big]\psi^+_{jm}
  \nonumber \\ &
  -i \sum_{\substack{j=2k+3/2 \\ k\geq0}} \sum_{m=-j}^j
  \bar\psi^-_{jm} \Big[ j+ \frac12 + \frac q2 - i r \s \Big]\bar \psi^-_{jm}\ .
\end{align}
As a consequence, the determinant for the fermion modes equals to
\begin{equation}
\mathrm{Det}_\psi = \prod_{k\geq 0}
\left( 2k+1-\frac{q}{2}+ir\sigma\right)^{4k+2}
\left(2k+\frac{q}{2}-ir\sigma\right)^{4k}\ .
\end{equation}
One can easily generalize the above results  for a chiral
multiplet of weight $w$ under $G$ by the replacement $\s \to w\cdot \s$.

Combining these two expressions, we find that the one-loop contribution
from a chiral multiplet in the representation $\mathbf{R}$ under
the gauge group $G$ is
\begin{equation}\label{even}
  Z_\text{1-loop}^{chiral} = \frac{\mathrm{Det}_\phi}{\mathrm{Det}_\psi} =
\prod_{w \in \mathbf{R}}\prod_{k\geq 0}
\frac{2k+1-\frac{q}{2}+ir w\cdot\sigma}{2k+\frac{q}{2}-ir w\cdot\sigma}\ .
\end{equation}
This can be regularized with Gamma function representation
\begin{equation}
\Gamma(a) = \lim_{n_{max}\rightarrow\infty}\frac{n_{max}!(n_{max})^a}{\prod_{n=0}^{n_{max}} (a+n)}\ ,
\end{equation}
where we should take care to introduce the UV cutoff $\Lambda$ via
$r\Lambda \simeq 2k_{max}$ since $(2k+\cdots)/r$ are the physical
eigenvalues. Then,
\begin{eqnarray}\nonumber
 Z_\text{1-loop}^{chiral}&=&\prod_{w \in \mathbf{R}}
 \lim_{k_{max}\rightarrow\infty}(k_{max})^{^{\frac{1}{2}-\frac{q}{2}+irw\cdot\sigma}}
\cdot
\frac{\Gamma\left(\frac{q}{4}-irw\cdot\sigma/2\right)}{
\Gamma\left(\frac12-\frac{q}{4}+irw\cdot\sigma/2\right)}
\\\nonumber
&=&\prod_{w \in \mathbf{R}}
e^{\left[\frac{1-q}{2}+irw\cdot\sigma\right]\log (r\Lambda/2)}
\cdot
\frac{\Gamma\left(\frac{q}{4}-irw\cdot\sigma/2\right)}{
\Gamma\left(\frac12-\frac{q}{4}+irw\cdot\sigma/2\right)}
\cdot\frac{\Gamma\left(-\frac{q}{4}+irw\cdot\sigma/2\right)}{
\Gamma\left(-\frac{q}{4}+irw\cdot\sigma/2\right)}
\\
&=& \prod_{w \in \mathbf{R}}\frac{1}{2\sqrt{2\pi}}\cdot
e^{\left[\frac{1-q}{2}+irw\cdot\sigma\right]\log (r\Lambda)}
\frac{\Gamma\left(\frac{q}{4}-\frac{irw\cdot\sigma}{2}\right)\cdot
\Gamma\left(-\frac{q}{4}+\frac{irw\cdot\sigma}{2}\right)}{
\Gamma\left(-\frac{q}{2}+irw\cdot\sigma\right)}\ , \label{evenfull}
\end{eqnarray}
where we used
\begin{equation}\label{dup}
\Gamma\left(\frac12+x\right)\Gamma(x)= 2^{1-2x}\sqrt\pi~\Gamma(2x)\ ,
\end{equation}
for the last equality.
The exponential factor which diverges when $\Lambda\rightarrow\infty$
is understood to be one-loop running of the FI-parameter
and appearance of central charge defined as $c\equiv 3(\sum_{i}(1-q_i)-d_G)$
when combined with vector multiplet contribution.

\paragraph{Odd Holonomy}
Let us now in turn consider the fluctuation near the second
saddle point with nontrivial holonomy.
At the odd holonomy fixed point, the boundary condition for charged
field must be twisted by $e^{iw\cdot h}=\pm 1$,
where $e^{ih\cdot H}$ is the $Z_2$ holonomy with unit-normalized Cartan generators $H$.
The chiral fields can then be classified into two classes, with even charge $w_e$
and with odd charge $w_o$, respectively, depending on the above sign. For
even ones, $w_e$, one-loop determinant is unchanged from the even holonomy
case, so we focus on a chiral multiplet with odd charge
$w_o$
%
\begin{align}
  \text{exp}\Big[ i \int_C w_o\cdot A \Big] = - 1\ .
\end{align}
Effectively, we impose the twisted projection condition
on those carrying odd charges $w_o$ as
\begin{align}
  \phi (\pi-\th,\pi+\varphi) = & - \phi(\th, \varphi)\ ,
  \nonumber \\
  \psi_\pm ( \pi -\th, \pi+\varphi) = & + i \psi_\mp (\th,\varphi) \ ,
  \nonumber \\
  \bar \psi_\pm (\pi-\th,\pi+\varphi) = & - i \bar \psi_\mp (\th,\varphi)\ ,
  \nonumber \\
  F (\pi -\th, \pi+\varphi) = & - F (\th,\varphi) \ ,
  \label{twistedchiralproj}
\end{align}
without a background gauge field. Thus the spectral analysis
is parallel to the previous one except
the twisted projection picks exactly opposite eigenvalues,
which were projected out under the original B-type parity action.
Therefore, one obtains
\begin{equation}
\mathrm{Det}_\phi = \left[\prod_{k\geq 0}
\left(2k+1+\frac{q}{2}-irw_o\cdot\sigma\right)^{4k+3}
\prod_{k\geq 1}\left(2k-\frac{q}{2}+irw_o\cdot\sigma\right)^{4k-1}\right]\ ,
\end{equation}
for bosons, and
\begin{equation}
\mathrm{Det}_\psi = \prod_{k\geq 0}
\left( 2k-\frac{q}{2}+ir w_o\cdot\sigma\right)^{4k}
\left(2k+1+\frac{q}{2}-ir w_o\cdot\sigma\right)^{4k+2}\ ,
\end{equation}
for fermions. Hence the one-loop determinant at this saddle point becomes
\begin{equation}\label{odd}
Z_\text{1-loop}^{chiral} =
\prod_{w_o\in\mathbf{R}}
\prod_{k\geq 0}
\frac{2k+2-\frac{q}{2}+irw_o\cdot\sigma}{2k+1+\frac{q}{2}-irw_o\cdot\sigma}\ .
\end{equation}
With the same procedure, we can further simplify this expression as
\begin{eqnarray} \label{oddfull}
Z_\text{1-loop}^{chiral}
&=&\prod_{w_o\in\mathbf{R}}\lim_{k_{max}\rightarrow\infty}(k_{max})^{\frac12-\frac{q}{2}+irw_o\cdot\sigma}
\frac{\Gamma\left(\frac12 +\frac{q}{4}-\frac{irw_o\cdot\sigma}{2}\right)}{
\Gamma\left(1-\frac{q}{4}+\frac{irw_o\cdot\sigma}{2}\right)}\\
&=&\prod_{w_o\in\mathbf{R}}2\sqrt{2\pi}\cdot e^{\left[\frac{1-q}{2}+irw_o\cdot\sigma\right]\log(r\Lambda)}
\frac{\Gamma\left(\frac{q}{2}-irw_o\cdot\sigma\right)}{
\Gamma\left(-\frac{q}{4}+\frac{irw_o\cdot\sigma}{2}\right)
\Gamma\left(\frac{q}{4}-\frac{irw_o\cdot\sigma}{2}\right)}
\cdot\frac{1}{-\frac{q}{2}+irw_0\cdot\sigma}\ .\nonumber
\end{eqnarray}

\subsection{Parity Accompanied by Flavor Rotations}\label{flavor}

For theories with non-trivial flavor symmetry, we can
enrich the $Z_2$ projection by combination
with flavor rotations, i.e.,
\begin{eqnarray}\nonumber
\phi^i(x) &\rightarrow& {M^i}_j\phi^j(x')\ ,\\
\psi^i_\pm(x) &\rightarrow& {M^i}_j\psi_\mp^j(x')\ ,
\end{eqnarray}
where ${M^i}_j$ is a flavor rotation which squares
to the identity. Let us consider the simplest example where
${M^i}_j$ exchanges two chiral multiplets
$\Phi^1(x) \leftrightarrow \Phi^2(x')$.
The contribution of these modes to the 1-loop
determinant is easily obtained, by noting that fluctuations
of one of $\Phi^{1,2}$ is completely determined by
that of the other in the opposite hemisphere.
Hence, these two effectively contribute as one chiral
multiplet without $Z_2$ projection, i.e.,
that of the full two-sphere partition function
\begin{equation}
\prod_{w\in \mathbf{R}} e^{[1-q+2irw\cdot\sigma]\log (r\Lambda)}
\cdot
\frac{\Gamma\left(\frac{q}{2}-irw\cdot\sigma\right)}{
\Gamma\left(1-\frac{q}{2}+irw\cdot\sigma\right)}\ ,
\end{equation}
calculated in Ref.~\cite{Doroud:2012xw,Benini:2012ui}.

All other $Z_2$ flavor transformations are generated
by combination of the above rotation and a gauge transformation.
For example, we can consider a projection of type $\Phi^1(x) \rightarrow -\Phi^1(x')$,
when the superpotential respects such symmetry.
The result of this sign flip is the same as in \eqref{odd}, so we find
\begin{equation}
\prod_{w\in \mathbf{R}}
2\sqrt{2\pi}\cdot e^{\left[\frac{1-q}{2}+irw\cdot\sigma\right]\log(r\Lambda)}
\cdot
\frac{\Gamma\left(\frac{q}{2}-irw\cdot\sigma\right)}{
\Gamma\left(-\frac{q}{4}+\frac{irw\cdot\sigma}{2}\right)
\Gamma\left(\frac{q}{4}-\frac{irw\cdot\sigma}{2}\right)}\cdot\frac{1}
{-\frac{q}{2}+irw\cdot\sigma}\ .
\end{equation}
These observations will be useful in the next section where
we consider lower-dimensional Orientifold planes embedded
as a hypersurface in the Calabi-Yau ambient space.

\subsection{Vector Multiplets}

Finally, we come to the vector multiplets.
We follow the Fadeev-Popov method to deal with the gauge symmetry,
and introduce ghost fields $c,\bar c$.
Up to the quadratic order, the action around the saddle point is
\begin{eqnarray}
  S_{vector} = S^b_{vector} + S^f_{vector} +S^{FP}_{vector}\ ,
\end{eqnarray}
where
\begin{align}
  S^b_{vec} = & \int \frac12 \text{Tr} \Big[ Da \wedge \ast Da
  - \big[ \s , a\big] \wedge \big[\s ,\ast a\big]
  + D \s_1 \wedge \ast D\s_1 - \big[ \s, \s_1\big]\wedge \big[\s,\ast \s_1\big]
  \nonumber \\ &
  + \frac{1}{r^2} \s_1 \wedge \ast \s_1
  + D \varphi \wedge \ast D\varphi +  \frac 2r Da \wedge \s_1 + i D\varphi \wedge \big[ \s, \ast a \big]
  + i \big[\s,a \big]\wedge \ast D\varphi \Big]\ ,
  \nonumber \\
  S^f_{vec} = & \int d^2x \sqrt{g} \  \frac 12 \text{Tr}  \Big[ \bar \l \g^3 \left(i \g^3 \g^i D_i \l
  + \big[\s, \l \big]\right) \Big] \ ,
  \nonumber \\
  S^{FP}_{vec} = & \int d^2x \sqrt{g} \ \text{Tr} \Big[ D_\m \bar c D_\m c
  + \frac12 f\wedge \ast f \Big]
  \label{vectorb}\ ,
\end{align}
with the gauge fixing functional
\begin{align}
  f = \ast D \ast a \ .
\end{align}
Here $a$ and $\varphi$ are the small fluctuation part  of the gauge field
and of the scalar field $\s_2$, respectively,
\begin{align}
  A = A_\text{flat} + a \ , \qquad \s_2 = \s + \varphi \ .
\end{align}
%

\paragraph{Even Holonomy}

When the holonomy is trivial, we impose the ordinary type B projection condition
\begin{align}
  A(\pi-\th,\pi + \varphi) = & + A(\th,\varphi) \ ,
  \nonumber \\
  \s_1(\pi-\th,\pi + \varphi) = & -\s_1(\th,\varphi)\ ,
  \nonumber \\
  \s_2(\pi-\th,\pi + \varphi) = & +\s_2(\th,\varphi)\ ,
  \nonumber \\
  \l_\pm (\pi-\th,\pi + \varphi) = & + i \l_\mp(\th,\varphi)\ ,
  \nonumber \\
  \bar \l_\pm (\pi-\th,\pi + \varphi) = & - i \bar \l_\mp(\th,\varphi)\ ,
  \nonumber \\
  D(\pi-\th,\pi + \varphi) = & +D(\th,\varphi)\ .
  \label{vectorproj}
\end{align}
First, decompose all the fluctuation fields into Cartan-Weyl basis, and
then consider the off-diagonal modes carrying the charge $\a$, a root of $G$.
In terms of the one-form and the scalar spherical
harmonics $C_{jm}^\lambda$\footnote{Useful properties of $C_{jm}^\lambda$ are summarized in
appendix \ref{harmonics}.}, $Y_{jm}$, one can expand the bosonic fluctuations $a^\a$, $\varphi^\a$,
and $\s_1^\a$ as
\begin{align}
a = & \sum_{\substack{j=2k\\k \geq 1}} \sum_{m=-j}^j
a_{jm}^1 C_{jm}^1 + \sum_{\substack{j=2k+1\\k \geq 0}} \sum_{m=-j}^j
a_{jm}^2 C_{jm}^2 \ ,
\nonumber \\
\s_1 = & \sum_{\substack{j=2k+1 \\ k \geq 0}}\sum_{m=-j}^j \s^1_{jm} Y_{jm}\ ,
\nonumber \\
\varphi = & \sum_{\substack{j=2k \\ k \geq0}}\sum_{m=-j}^j \varphi_{jm} Y_{jm}\ ,
\end{align}
under the projection condition (\ref{vectorproj}).
From now on, the superscript $\a$ is suppressed unless it causes any confusion.
The Laplacian operator $\CO_b^{(1)}$ acting on $(a^2_{jm},\s^1_{jm})$
can be summarized into
\begin{align}
  \CO_b^{(1)} \doteq \begin{pmatrix}
    j(j+1) + (\s\cdot\a)^2 & \sqrt{j(j+1)} \\
    \sqrt{j(j+1)} & j(j+1) + (\s\cdot\a)^2 + 1
  \end{pmatrix}\ ,
\end{align}
with $j=2k+1$ ($k\geq 0$). The determinant of this operator is therefore,
\begin{align}
  \sqrt{\det\CO_b^{(1)}} = &\prod_{k\geq0} \Big[ (2k+1)(2k+2) \Big]^{(4k+3)r_G}
  \nonumber \\ & \times
   \prod_{\a\in \D_+} \prod_{k\geq0}
   \Big[ \left( (2k+1)^2 + (\a\cdot \s)^2 \right)
  \left((2k+2)^2 + (\a\cdot\s)^2 \right) \Big]^{4k+3}\ ,
\end{align}
where $r_G$ is rank of the gauge group. The operator $\CO_b^{(2)}$
acting on the modes $(a^1_{jm},\varphi_{jm})$ with $j=2k$ ($k\geq1$)
can be read from (\ref{vectorb}),
\begin{align}
  \CO_b^{(2)} \doteq \begin{pmatrix}
  j(j+1) + (\s\cdot \a)^2 & i \sqrt{j(j+1)}(\s\cdot \a) \\
   - i \sqrt{j(j+1)}(\s\cdot \a) & j ( j+1)
  \end{pmatrix}\ .
\end{align}
When $j=0$, the operator has a vanishing eigenvalue that
corresponds to the shift of the saddle point $\s_2=\s$. The determinant of
this operator is therefore
\begin{align}
  \sqrt{\det{}' \CO_b^{(2)}} =
  \prod_{k=1} \Big[ 2k(2k+1) \Big]^{(4k+1)d_G}  \ ,
\end{align}
where $d_G$ is dimension of the gauge group $G$, and the prime in $\det{}'$
denotes the fact that the zero mode of $\sigma_2$ is removed.
For the ghosts, we require the same projection condition as $\varphi,\bar\varphi$,
and find
\begin{equation}
\det\CO_{FP}=\prod_{k=1} \Big[2k(2k+1)\Big]^{d_G(4j+1)} \ ,
\end{equation}
which cancels with $\CO_{b}^{(2)}$ determinant exactly.
For fermions, the structure of determinants are essentially the same as
that of the adjoint chiral multiplet with the twisted projection condition.
Therefore, gaugino with root $\alpha$ contributes
\begin{align}
\det \CO_{\lambda}= &
\prod_{k \geq 0} \Big[(2k+1)(2k+2) \Big]^{r_G(4k+2)}
\nonumber \\ & \times
\prod_{\alpha\in \D_+}\prod_{k \geq 0} \Big[ \left( (2k+1)^2 + (\alpha\cdot\sigma)^2 \right)
\left( (2k+2)^2 + (\a\cdot\s)^2 \right) \Big]^{4k+2}\ .
\label{gaugino}
\end{align}

Let us combine all these contributions from vector multiplets
together. The Cartan part of the vector multiplets contributes,
\begin{equation}
\prod_{j=0}\left(\frac{2j+2}{2j+3}\right)^{r_G} =
\left[\frac{\Gamma\left(\frac{3}{2}\right)}{\Gamma(1)}
\cdot e^{-\frac12\log(r\Lambda/2)}\right]^{r_G} =
\left(\frac{\pi}{2}\right)^{\frac{r_G}{2}}  e^{-\frac{r_G}{2}\log(r\Lambda)}\ .
\end{equation}
while the ``off-diagonal part" regularize to
\begin{eqnarray}\nonumber
&&\prod_{\alpha\in\D_+}\prod_{k=0}
\frac{(2k)^2+ (\alpha\cdot\sigma)^2 }{(2k+1)^2 + (\alpha\cdot\sigma)^2}
\cdot\frac{1}{(\alpha\cdot\sigma)^2}\cr
&=&
e^{-\frac12 (d_G-r_G)\log(r\Lambda/2)}
\prod_{\alpha\in \D_+}
\frac{\Gamma\left(\frac12+\frac{i\alpha\cdot\sigma}{2}\right)
\Gamma\left(\frac12-\frac{i\alpha\cdot\sigma}{2}\right)}{
4\cdot\Gamma\left(1+\frac{i\alpha\cdot\sigma}{2}\right)
\Gamma\left(1-\frac{i\alpha\cdot\sigma}{2}\right)}
\\\nonumber
&=&
 e^{-\frac12 (d_G-r_G)\log(r\Lambda/2)}
\prod_{\alpha\in \D_+}
\frac{2\pi\sin\left[\frac{\pi\alpha\cdot\sigma}{2}\right]}{\sin\pi\alpha\cdot\sigma}
\cdot\frac{\sin\left[\frac{\pi\alpha\cdot\sigma}{2}\right]}{2\pi\alpha\cdot\sigma}
\\
&=&
 e^{-\frac12 (d_G-r_G)\log(r\Lambda)}
\prod_{\alpha\in\D_+}
\frac{1}{\alpha\cdot\sigma}\cdot\tan\left(\frac{\pi\alpha\cdot\sigma}{2}\right)\ .
\end{eqnarray}
As the zero mode part contributes
\begin{equation}
\frac{1}{|W_G|}\int d^{r_G}\sigma\prod_{\alpha\cdot\sigma>0}(\alpha\cdot\sigma)^2\ ,
\end{equation}
with the Vandermonde determinant and the Weyl factor,
we obtain the even holonomy part of the partition function,
where the vector multiplet contributions in the even holonomy sector
can be displayed explicitly as
\begin{eqnarray}\label{vector1}
Z^{even}&=&\frac{1}{|W_G|}\int d^{r_G}\sigma
\left(\frac{\pi}{2}\right)^{\frac{r_G}{2}}
\cdot e^{-\frac{d_G}{2}\log(r\Lambda)}\\
&&\hskip 1cm \times \prod_{\alpha\in \Delta_+}
\alpha\cdot\sigma\tan\left(\frac{\pi\alpha\cdot\sigma}{2}\right) \times \cdots\ ,
\end{eqnarray}
where the ellipsis reminds us that for the GLSM partition function, we
need to insert, multiplicatively, the 1-loop contributions from
the chiral multiplets in the integrand.

\paragraph{Odd Holonomy}
At the odd holonomy fixed point, the boundary condition for the
vector multiplet fluctuation must be twisted by $e^{i\alpha\cdot h}=\pm 1$,
where, as before, $e^{ih\cdot H}$ is the $Z_2$ holonomy with the Cartan
generators $H$. Thus, we only need to modify, in Eq.~(\ref{vector1}), as
\begin{equation}
\tan\left(\frac{\pi\alpha\cdot\sigma}{2}\right)\quad\rightarrow\quad
\cot\left(\frac{\pi\alpha\cdot\sigma}{2}\right)\ ,
\end{equation}
for each and every root with $e^{i\alpha\cdot h}=-1$. So, splitting the
positive root space $\Delta_+$ into the even part $\Delta_+^e$ and the
odd part $\Delta_+^e$, relative to the holonomy $e^{ih\cdot H}$, we find
that the odd holonomy sector contributes additively to the partition function
\begin{eqnarray}\label{vector2}
Z^{odd}&=&\frac{\eta}{|W_G|}\int d^{r_G}\sigma
\left(\frac{\pi}{2}\right)^{\frac{r_G}{2}}
\cdot e^{-\frac{d_G}{2}\log(r\Lambda)}\\\cr
&&\hskip 1cm\times
\prod_{\alpha_e\in \Delta_+^e}
\alpha_e\cdot\sigma\tan\left(\frac{\pi\alpha_e\cdot\sigma}{2}\right)\prod_{\alpha_o\in \Delta_+^o}
\alpha_o\cdot\sigma\cot\left(\frac{\pi\alpha_o\cdot\sigma}{2}\right) \times \cdots\ ,\nonumber
\end{eqnarray}
where, again, the ellipsis in the integrand denotes multiplicative
contributions from the chiral multiplet 1-loop determinants.

The numerical factor $\eta=\pm 1$ represents our ignorance
regarding fermion determinants. As with any determinant computation
involving fermions, the signs of various 1-loop factors are difficult
to fix. Among such, $\eta$ which is the relative sign between the two
additive contributions, from the even holonomy and
the odd holonomy sectors, is an important physical quantity but is
not accessible from the Coulomb-phase GLSM computation.
For this reason, and also as a consistency check, we make a short excursion
to the mirror LG computation for the Abelian GLSM, in next section, which will
teach about how this relative sign may be fixed.

\section{Landau-Ginzburg Model and Mirror Symmetry}

Before we consider examples and the large volume limit, let us make a brief
look at the mirror pair of the Abelian GLSM. In particular, we consider
$U(1)$ theory with chiral multiplets $\Phi_a$ with gauge charges $Q_a$.
As shown by Hori and Vafa \cite{Hori:2000kt}, the mirror theory is
a Landau-Ginzburg (LG) type with twisted chiral multiplet $Y_a$'s
and the twisted superpotential $W(Y_a)$, generated by the vortex instantons.
On $\mathbb{RP}^2$, the supersymmetric Lagrangian of a LG model with twisted
chiral multiplets takes the following form
\begin{align}
  \CL = \CL_{twisted} + \CL_W\ ,
\end{align}
with
\begin{align}
  \CL_{twisted} = D^\m \bar Y D_\m Y
  + i \bar \chi \g^m D_m \chi
  +  \bar G G \ ,
\end{align}
and the twisted superpotential terms,
\begin{align}
  \CL_W  = & + \Big[ - i W'(Y) G
  - W''(Y) \bar \chi \g_- \chi + \frac ir W(Y) \Big]
  \nonumber \\ &
  + \Big[ - i \bar W'(\bar Y)  \bar G
  + \bar W''(\bar Y) \bar \chi \g_+  \chi + \frac ir \bar W (\bar Y) \Big] \ ,
\end{align}
where $\g_\pm = \frac{1+\g^3}{2}$.
One can show that the above Lagrangian is invariant under
the supersymmetric variation rules given by
\begin{align}
  \d Y = & + i  \bar \e \g_- \chi  - i  \e  \g_+ \bar \chi  \ , \nonumber
  \\
  \d \overline Y = & - i \bar \e \g_+ \chi + i\e \g_- \bar \chi \ ,
  \nonumber \\
  \d \chi = & + \g^\m \g_+ \e D_\m Y - \g^\m \g_- \e D_\m \overline Y
  - \g_+ \e \bar G  - \g_- \e G  \ ,
  \nonumber \\
  \d \bar \chi = &  +  \g^\m \g_+ \bar \e D_\m \overline Y
  - \g^\m \g_- \bar \e D_\m Y
  + \g_+ \bar \e G
  + \g_- \bar \e \bar G \ ,
  \nonumber \\
  \d G = & - i\bar \e \g^\m \g_- D_\m \chi + i \e \g^\m \g_+ D_\m \bar \chi \ ,
  \nonumber \\
  \d \bar G = & - i \bar \e \g^\m \g_+ D_\m \chi + i \e \g^\m \g_- D_\m \bar \chi \ ,
  \label{tSUSY}
\end{align}
where $\epsilon$ and $\bar\epsilon$ are the Killing spinors (\ref{killing}).
The kinetic terms are again Q-exact \cite{Gomis:2012wy, Doroud:2013pka},
\begin{align}
  \CL_\text{twisted} =\delta_\epsilon \delta_{\bar \epsilon} \Big[ \frac ir \bar Y Y - i \bar G Y - i \bar Y G \Big] \ .
\end{align}

Type B-parity action on the twisted chiral fields resembles the type A-parity on
the chiral fields, naturally, which we first outline. One important fact,
perhaps not too obvious immediately, is that the parity action which
flips $Y$ to $\bar Y$ should be accompanied by a half-shift of the imaginary
part, in order to preserve the action. Due to this, the fixed submanifolds
are spanned by
\begin{equation}
Y=x +in \frac{\pi}{2}\ ,
\end{equation}
with $n=\pm1$.

On this mirror side, the role of $\theta$ angle becomes more visible.
{}From the equation of motion for the vector multiplet, we learn allowed
values of $n$'s have to be such that
\begin{align}
  \frac12\sum_a Q_a n^a = \frac{\theta}{\pi} \quad {\rm mod}\  Z_2\ ,
\end{align}
which restricts the sum over $n^a=\pm 1$ into two disjoint sets, depending
on the value of $\theta$. Recall that the GLSM localization procedure
was unable to see the distinction between these two values. Instead, one
finds ambiguity in the sign of the determiants, especially,
relative sign between different holonomy sectors. The two such sectors are
topologically distinct, so one can introduce this relative sign as a
parameter of the theory, which we called $\eta$ in  \eqref{vector2},
which will be presently related to $\theta$.

\subsection{Parity on the Mirror}

Under the type B-parity (\ref{bparitykilling}), one can show that the projection conditions
are
\begin{align}
  Y( \pi - \th, \pi + \varphi) = \bar Y (\th,\varphi) + \text{constant}\ ,
  \label{unfixed}
\end{align}
and
\begin{align}
  \chi_\pm ( \pi - \th, \pi + \varphi) =  & + i \chi_\mp(\th,\varphi) \ ,
  \nonumber \\
  \bar \chi_\pm ( \pi - \th, \pi + \varphi) = &  - i \bar \chi_\mp (\th, \varphi)\ ,
  \nonumber \\
  G ( \pi -\th, \pi+\varphi) = & + \bar G (\th,\varphi) \
  \label{twistedproj1}
\end{align}
are consistent to the SUSY variation rules, for free theories.
In order to fix the constant term in (\ref{unfixed}), we need to consider interactions
such as twisted superpotential terms.

First, recall that the gauge multiplet
can be written as a twisted chiral $\Sigma$, where
\begin{gather}
  Y = \s_2 + i\s_1\ , \qquad
  G = D  + i \left( F_{12} + \frac{\s_1}{r} \right) \ ,
  \nonumber \\
  \chi =\l \ , \qquad \bar \chi = \bar \l\ .
\end{gather}
As discussed above, we impose the projection conditions
  \begin{align}
    \s_1 ( \pi - \th , \pi+\varphi) = & - \s_1 (\th,\varphi) \ ,
    \nonumber \\
    \s_2 (\pi-\th,\pi+\varphi) = & + \s_2 (\th,\varphi)\ ,
  \end{align}
in order to introduce a minimal coupling of a charged chiral multiplet.
It implies that
\begin{align}
    \Sigma(\pi-\th, \pi +\varphi) = \bar \Sigma (\th, \varphi) \ .
    \label{twistedproj2}
\end{align}
%
Note also that $\Sigma$ enters the tree-level twisted superpotential linearly as
\begin{align}
  W = - \frac i2 \t \Sigma\ ,
\end{align}
with $\t= i \xi + \frac{\th}{2\pi}$, which leads to the FI coupling and 2d topological
term
\begin{align}
    \CL_W+\CL_{\bar W} = - i \xi \left(D -\frac{\s_2}{r} \right) - i \frac{\th}{2\pi}F_{12}\ .
\end{align}
  Note that the complexified FI parameter is periodic $\t \simeq \t + n$ ($n \in \mathbb{Z}$).
  In order to make the interaction invariant under the type B Orientifold action,
  the parameter $\t$ has to satisfy the following condition,
  \begin{align}
    \t + \bar\t = n\ ,  \qquad n \in \mathbb{Z}\ .
  \end{align}
  In other words, the allowed value for the two-dimensional theta angle is either
  \begin{align}
    \th = 0 ~\text{  or } \ \pi\ .
  \end{align}

Second, let us consider a simple example mirror to the $U(1)$
GLSM with $n$ chiral multiplets of gauge charge $Q_a$ where $a$ runs from $1$ to $n$.
The chiral multiplets also carry $U(1)_V$ $R$-charges $q^a$ so that
the superpotential $\CW$ carries the $R$-charge two.
The mirror Landau-Ginzburg model involves $n$ neutral
twisted chiral multiplets $Y^a$ with period $2\pi i$.
The dual description also comes with the
following twisted superpotential
  \begin{align}
    W = - \frac{1}{4\pi} \Big[ \S \left( \sum_{a=1}^n Q_a Y^a + 2\pi i \t\right)
    + \frac ir \sum_{a=1}^n e^{-Y^a} \Big]\ .
    \label{tpotential}
  \end{align}
  %
At low-energy, the field-strength  multiplet $\S$
is effectively a Lagrange multiplier, leading to the constraint:
\begin{align}
  \sum_{a=1}^n Q_a Y^a = -2\pi i \t\ .
\end{align}
To make these Toda-like interaction terms invariant under the type B-parity,
one has to fix the constant piece in (\ref{unfixed}) by $i\pi$. That is,
\begin{align}
    Y (\pi -\th,\pi+\varphi) = \bar Y(\th,\varphi) + i\pi \ .
\label{twistedproj3}
\end{align}
%

%

\subsection{Partition Function on $\mathbb{RP}^2$}

Choosing the kinetic terms $\CL_{twisted}$ as Q-exact deformation terms,
one can show that the path-integral localizes onto
\begin{align}
  Y = x + i y\ ,
\end{align}
where $x$ and $y$ are real constants \cite{Gomis:2012wy}.
To obey the projection conditions (\ref{twistedproj2}) and
(\ref{twistedproj3}), the supersymmetric saddle points are
\begin{align}
  \s_2 = \sigma \ , \qquad \s_1 = 0 \ , \qquad F_{12} = 0\ ,
\end{align}
and
\begin{align}
  Y^a = x^a + \frac {i\pi}{2} n^a  \ ,
\end{align}
where $x^a$ and $\sigma$ are real constants over $\mathbb{RP}^2$.
Here $n^a = \pm 1$ obeying the constraint, for $\theta=0$,
\begin{align}
  \frac12\sum_a Q_a n^a = 2 m \ , \qquad m \in \mathbb{Z} \ ,
\end{align}
and for $\theta=\pi$,
\begin{align}
  \frac12\sum_a Q_a n^a = 2 m +1\ , \qquad m \in \mathbb{Z}\ ,
\end{align}
obeying the constraint
\begin{align}
  \sum_a Q_a Y^a = - 2\pi i \t\ .
\end{align}

\paragraph{$\mathbb{RP}^2$ Partition Function}

It is easy to show that  one-loop determinants around the above supersymmetric
saddle points are trivial in a sense that they are independent of $\s$ and $x^a$.
One can show that the partition function of the mirror LG model with the twisted
superpotential (\ref{tpotential}) on $\mathbb{RP}^2$ reduces to an ordinary contour
integral,\footnote{
We used for the last equality an integral formula
\begin{align}\nonumber
  \int_{-\infty}^{\infty} dx \ e^{i p x} \cos\left[ e^{-x} + z\right]
  = \cos\left[ \frac {i\pi p}{2} - z\right] \G[ - i p]\ ,
  \text{ if } - 1 < \mathfrak{Re}[ip] < 0\ .
\end{align}}
\begin{align} \label{LG}
  Z_\text{LG} \simeq & \ \int_{-\infty}^\infty d\s \prod_a
  \left[ \int_{-\infty}^\infty dx^a e^{-\frac{q_a}{2}x^a} \right] \
  \sum_{n^a = \pm 1}  \frac12 \left( 1 \pm e^{i\pi  Q_a n^a/2} \right) \cdot
  e^{ir\s (  Q_a x^a -2\pi \xi) } \cdot
  e^{i e^{-x^a} \sin (\pi n_a/2)}
  \nonumber \\
  = & \  \int_{-\infty}^\infty d\s e^{ -2i\pi r\sigma \xi }\bigg\{
  \prod_a \cos\Big[\frac{\pi}{2} \left( \frac{q_a}{2} - ir Q_a \s\right)\Big] \G\Big[ \frac{q_a}{2} - ir Q_a \s \Big]
  \nonumber \\
 &  \qquad\qquad\qquad\quad\pm
 \ \prod_b \cos\Big[\frac{\pi}{2} \left( \frac{q_b}{2} - ir Q_b \s\right)
 -\frac{\pi}{2}Q_b\Big] \G\Big[ \frac{q_b}{2} - ir Q_b \s \Big]\bigg\}\ ,
\end{align}
%
where  ``$\simeq$" symbol in the first line reflects our ignorance of the overall
numerical normalization of the integration measure.
Here the factors $e^{-\frac{q_a}{2}x^a}$ reflect the important
fact that the proper variables describing the mirror LG
model are $X^a = e^{-\frac{q_a}{2}Y^a}$ rather than $Y^a$ \cite{Gomis:2012wy}.
Below, we  compare to the GLSM
side up to this normalization issue. The signs $\pm$ are for $\theta=0$ and $\theta=\pi$ respectively.

The parity projection that leads to Eq.~(\ref{LG}) assumes no specific
flavor symmetry in the original GLSM, and thus must be the mirror of the
spacetime-filling case of section 3.2. In the trivial holonomy sector,
we start with the last line of Eq.~(\ref{evenfull}) and
%
%
use the identities
\begin{align}
&\G\left( \frac12 + x\right) \G( x) = 2^{1-2x} \sqrt\pi  \G(2x)\ ,
\qquad
\G\big(1-x\big) \G\big(x\big) = \frac{\pi}{\sin{\pi x} }\ ,
\end{align}
to massage the one-loop determinant into
\begin{align}
  Z_\text{1-loop}^\text{trivial} =
  \G\Big[ \frac q2 - i Q \s\Big]
  \cos\left[ \frac{\pi}{2} \left( \frac{q}{2} - i Q \s\right) \right] \times \sqrt{\frac{2}{\pi}}
  e^{\left[\frac{1-q}{2} + i Q \s\right] \log(r \L) }\ .
\end{align}
In the nontrivial holonomy, a chiral multiplet carrying the
even charge $Q_a=Q_e$, the same result holds,
\begin{align}
  Z_{\text{1-loop}, Q_e}^\text{nontrivial} =
  \G\left[ \frac q2 - i Q_e \s\right]
  \cos\left[  \frac{\pi}{2} \left( \frac q2 - i Q_e\s\right) \right]  \times \sqrt{\frac{2}{\pi}}
  e^{\left[\frac{1-q}{2} + i Q \s\right] \log(r \L) }\ ,
\end{align}
while for the odd charge, $Q_a=Q_o$, the partition function becomes
\begin{align}
  Z_{\text{1-loop},Q_o}^\text{nontrivial} = & \
  \prod_{k\geq0}  \frac{2k+2 - \frac q2 + i Q_o \s}{2k+1 - \frac q2 + i Q_o \s}
  \nonumber \\ = & \
   \frac{\G\big( \frac12 + \frac q4 - \frac{i Q_o \s}{2}\big)}
  {\G\big(1 - \frac q4 + \frac{i Q_o \s}{2}\big)} \times
  e^{\left[\frac{1-q}{2} + i Q_o\s\right] \log(r\L/2) }
  \nonumber \\ = & \
   \G \left[ \frac q2 - i Q_o \s \right]
  \sin\left[  \frac{\pi}{2} \left( \frac  q2 - i Q_o \s \right) \right]
   \times \sqrt{\frac{2}{\pi}}
  e^{\left[\frac{1-q}{2} + i Q \s\right] \log( r \L) }\ .
\end{align}
Thus one can conclude that the first term in the final expression (\ref{LG})
of the LG partition function corresponds to the partition function of GLSM
with the even holonomy, while the second term corresponds to the partition
function with the odd holonomy.

After interpreting the exponentiated log piece as the renormalization
of $\xi$, we learn two additional facts. First, the common overall
normalization $\sqrt{2/\pi}$ should be incorporated into the
measure on the mirror LG side. Second, an additional relative sign
$\eta\equiv \pm \prod_a (-1)^{[Q_a/2]}$ (for $\theta=0,\pi$,
respectively) should sit between the trivial and the nontrivial
holonomy contributions in the GLSM side, and tells us how the
discrete $\theta$ angles must be understood from the localization
computation: It dictates how the contributions from topologically
distinct holonomy sectors should be summed. When the Orientifold projection
produces more than one Orientifold planes, which we will see in examples
of next section, this sign $\eta$ also distinguishes relative $O^\pm$
type of these Orientifold planes.\footnote{Recall that $\tilde O^\pm$
type Orientifolds involve turning on discrete RR-flux \cite{Witten:1998xy},
and thus are not accessible from GLSM. See also Ref.~\cite{Gao:2010ava}
for relationship between $\theta$ angle and
Orientifold plane type for various dimensions.}

\section{Orientifolds in Calabi-Yau Hypersurface}\label{CY}

In this section, we consider the Orientifolds for a prototype Calabi-Yau
manifold $\CX$, i.e., a degree $N$ hypersurface of ${CP}^{N-1}$.
At the level of GLSM, the chiral field contents are
\begin{equation}\begin{array}{c|cc}
&U(1)_G&U(1)_V\\
\hline
X_{i=1,\cdots N} & 1 &q\\
P& -N &2-Nq
\end{array}
\end{equation}
where we displayed the gauge and the vector $R$-charges. As usual, the
superpotential takes the form $P\cdot G_N(X)$ with degree $N$ homogeneous polynomial
$G_N$. For simplicity, we will call $\epsilon=q/2-ir\sigma$ below, and
assume $N$ odd. For $N=$ even, the $P$ multiplet contributions
from even and odd holonomy are exchanged. The number $q$ is in principle
arbitrary as it can be shifted by mixing $U(1)_G$ and $U(1)_V$,
but we restrict it to be in the range $0<q<2/N$ \cite{Gomis:2012wy}.

The main goal of this section is to extract the large volume expressions
for the central charges of  Orientifold planes. Traditionally, the latter
were expressed in terms of the $\sqrt{\CL}$ class, but just as with D-brane
central charge, we will see that $\hat \Gamma_c$ class
enters and corrects the expression.
$\hat\Gamma_c$ is a multiplicative class associated with the function
\cite{Hori:2013ika,Libgober,Iritani,Katzarkov:2008hs,Halverson:2013qca}
\begin{equation}
\Gamma\left(1+\frac{x}{2\pi i}\right) \ ,
\end{equation}
so that, for any holomorphic bundle $\CF$, an important identity
\begin{equation}
\hat\Gamma_c (\CF)\hat\Gamma_c(-\CF)= \CA(\CF)
\end{equation}
holds. In terms of the Chern characters, it can be expanded as
\begin{eqnarray}
\hat\Gamma_c({\CF}) 
&=& \exp \left[\frac{i\gamma}{2\pi}ch_1(\CF)+\sum_{k\geq2}
\left(\frac{i}{2\pi}\right)^k(k-1)!\zeta(k)ch_k(\CF)\right]\ ,
\end{eqnarray}
where $\gamma=0.577...$ is the Euler-Mascheroni constant,
and $\zeta(k)$ is the Riemann zeta function.

The results
from this hypersurface examples suggest that, for a general Orientifold plane
that wraps a cycle $\CM $ in the Calabi-Yau $\CX$,
with the tangent bundle $\CT\CM $ and the normal bundle $\CN\CM$ with respect to $\CX$, we must
correct the characteristic class that appear in the central charge as
\begin{equation}
\frac{\sqrt{\CL(\CT\CM/4)}}{\sqrt{\CL(\CN\CM/4)}}\quad\rightarrow\quad
\frac{\CA(\CT\CM/2)}{\hat\Gamma_c(-\CT\CM)}\wedge\frac{\hat\Gamma_c(\CN\CM)}{\CA(\CN\CM/2)} \ .
\end{equation}
We devote the rest of this section  to derivation of this,
by isolating the perturbative contributions for Orientifolds wrapping (partially)
Calabi-Yau hypersurfaces in $\mathbb{CP}^{N-1}$.

\subsection{Spacetime-Filling Orientifolds}

First, let us consider the case where the Orientifold plane wraps $\CX$ entirely,
i.e., no flavor symmetry action is mixed with the B-parity projection.
With the classical contribution
\begin{equation}
Z_{classical}=e^{-i2\pi r\xi \sigma} = e^{-2\pi\xi(q/2-\epsilon)} ,
\end{equation}
we find
\begin{eqnarray}\nonumber
Z_{\mathbb{RP}^2} &=&
\int_{q/2-i\infty}^{q/2+i\infty}\frac{d\epsilon}{2\pi i}~~ \Bigg(
\beta_1\cdot
e^{2\pi\xi\epsilon}\left[\frac{\Gamma\left(\frac{\epsilon}{2}\right)
\Gamma\left(-\frac{\epsilon}{2}\right)}{\Gamma(-\epsilon)}\right]^N
\cdot\left[\frac{\Gamma\left(\frac{1-N\epsilon}{2}\right)
\Gamma\left(-\frac{1-N\epsilon}{2}\right)}{\Gamma(-1+N\epsilon)}\right]\\\label{CYh}
&&\hskip1cm +\eta\cdot  \beta_2\cdot e^{2\pi\xi\epsilon}
\left[\frac{\Gamma(\epsilon)/\epsilon}{\Gamma\left(\frac{\epsilon}{2}\right)
\Gamma\left(-\frac{\epsilon}{2}\right)}\right]^N\cdot
\left[\frac{\Gamma(1-N\epsilon)/(1-N\epsilon)}{\Gamma\left(\frac{1-N\epsilon}{2}\right)
\Gamma\left(-\frac{1-N\epsilon}{2}\right)}\right]\Bigg)\ ,\label{filling}
\end{eqnarray}
where the constants $\beta_{1,2}$ are
\begin{eqnarray}\nonumber
\beta_1&=& e^{-\pi\xi q}\cdot (2\pi)^{-N/2+1}\cdot2^{-(N+2)}\cdot e^{\frac{c}{6}\cdot\log(r\Lambda)}\ ,\\
\beta_2&=& e^{-\pi\xi q}\cdot (2\pi)^{N/2+2}\cdot2^N \cdot e^{\frac{c}{6}\log(r\Lambda)}\ ,
\end{eqnarray}
with $\eta = \pm \prod_a (-1)^{[Q_a/2]}$ (for $\theta=0,\pi$, respectively).
Strictly speaking, there is also an overall sign ambiguity, which together with
$\eta$ affect $O^\pm$ type of Orientifolds that reside in the each holonomy sector.
Recall that the two lines are, respectively, contributions from the even and
the odd holonomy sector.
Another common factor in $\beta_{1,2}$, $e^{\frac{c}{6}\log(r\Lambda)}$,
renormalizes the partition function. Because $\CX$ is Calabi-Yau,
$\xi$ is not renormalized but the partition function itself is
multiplicatively renormalized with the exponent $c/6 =(N-2)/2$ for this model.

The first factor in $\beta_{1,2}$, i.e., $ e^{-\pi\xi q}$, with an explicit
dependence on the  $R$-charge assignment, looks a little strange as $q$ is
not uniquely defined. Note that $q\rightarrow q+\delta$ is a shift
of $R$-charges by the gauge charges. Something similar happens for
hemisphere and also for $S^2$, where, for the latter, the partition function
having been identified with $e^{-K}$, the shift is understood to be a K\"ahler
transformation. On the other hand, $S^2$ partition function can be built
from a pair of hemisphere partition functions and a cylinder, so it is
to be expected that the hemisphere partition function should be a section
rather than a function. Along the same line of thinking, then, the crosscap
amplitudes should be no different from boundary state
amplitudes.\footnote{We are indebted to Kentaro Hori for explaining
this point to us.}  With this in mind, we choose to set $q\rightarrow 0^+$
from this point on as the canonical choice, following Ref.~\cite{Hori:2013ika}.
Note that the integral converges only when $q$ is positive real \cite{Gomis:2012wy}.

When $\xi>0$, the GLSM flows to the geometric phase in IR and
we should close the contour to the left infinity. For the
even holonomy sector, the relevant poles are those of $\Gamma(\epsilon/2)$
at $\epsilon=-2k~(k=0,1,2,\cdots)$.  For the odd holonomy sector,
the  relevant poles are those of $\Gamma(\epsilon)/
\Gamma(\epsilon/2)$ at $\epsilon=-(2k+1)~(k=0,1,2,\cdots)$. Poles of
other factors either cancel out among themselves or are located
outside of the contour.
Of these,  poles at $\e<0$ capture the world-sheet instanton
contributions, which are suppressed exponentially in the large volume
limit $\xi\gg 1$.

The perturbative part of the partition function,
appropriate for the large volume limit, comes entirely
from the pole at $\epsilon=0$. With \eqref{filling}, therefore, only
the even holonomy sector contributes, giving us
\begin{eqnarray}
Z_{\mathbb{RP}^2}^{pert.}&=&\beta_1\oint_{\epsilon=0}
\frac{d\epsilon}{2\pi i}~~
 e^{2\pi\xi\epsilon}\left[\frac{\Gamma\left(\frac{\epsilon}{2}\right)
\Gamma\left(-\frac{\epsilon}{2}\right)}{\Gamma(-\epsilon)}\right]^N
\cdot\left[\frac{\Gamma\left(\frac{1-N\epsilon}{2}\right)
\Gamma\left(-\frac{1-N\epsilon}{2}\right)}{\Gamma(-1+N\epsilon)}\right]\ .
\end{eqnarray}
We first invoke the identity
\begin{eqnarray}\label{identity1}
\Gamma\left(\frac12+x\right)\Gamma(x)= 2^{1-2x}\sqrt\pi~\Gamma(2x)\ .
\end{eqnarray}
to rewrite this as
\begin{eqnarray}
Z_{\mathbb{RP}^2}^{pert.}&=&8\pi \beta_1 \oint_{\epsilon=0}  \frac{d\epsilon}{2\pi i}~
 e^{2\pi\xi\epsilon}\left[\frac{\Gamma\left(\frac{\epsilon}{2}\right)
\Gamma\left(-\frac{\epsilon}{2}\right)}{\Gamma(-\epsilon)}\right]^N
\Bigg/\left[\frac{\Gamma\left(\frac{N\epsilon}{2}\right)
\Gamma\left(-\frac{N\epsilon}{2}\right)}{\Gamma(-N\epsilon)}\right]\\\nonumber
&=&8\pi \beta_1 \cdot 2^{2(N-1)}\oint_{\epsilon=0} \frac{d\epsilon}{2\pi i}~
 e^{2\pi\xi\epsilon}
\frac{N}{\epsilon^{N-1}} \cr&&\qquad\qquad\qquad \times
\left[\frac{\Gamma\left(1+\frac{\epsilon}{2}\right)
\Gamma\left(1-\frac{\epsilon}{2}\right)}{\Gamma(1-\epsilon)}\right]^N
\Biggl/\left[\frac{\Gamma\left(1+\frac{N\epsilon}{2}\right)
\Gamma\left(1-\frac{N\epsilon}{2}\right)}{\Gamma(1-N\epsilon)}\right]\ .
\end{eqnarray}
This can be further  rewritten as an integral over $\CX$, with
$H$ the hyperplane class of $\mathbb{CP}^{N-1}$,
\begin{eqnarray}
Z_{\mathbb{RP}^2}^{pert.} =C_0 \int_\CX
e^{-i\xi H}
\left[\frac{\Gamma\left(1+\frac{H}{4\pi i}\right)
\Gamma\left(1-\frac{H}{4\pi i}\right)}{\Gamma(1-\frac{H}{2\pi i})}\right]^N
\Bigg/\left[\frac{\Gamma\left(1+\frac{NH}{4\pi i}\right)
\Gamma\left(1-\frac{NH}{4\pi i}\right)}{\Gamma(1-\frac{NH}{2\pi i})}\right]\ ,
\end{eqnarray}
with $C_0=i^{N-2}(2\pi)^{N/2}2^{N-2}(\Lambda r)^{c/6}$. We used
 $\int_\CX H^{N-2}=N\int_{\mathbb{CP}^{N-1}}H^{N-1}=N$.

Since $\CX$ is a Calabi-Yau hypersurface  embedded in
$\mathbb{CP}^{N-1}$, we may also write
\begin{equation}
\hat\Gamma_c(\CT \CX) = \frac{\hat\Gamma_c (\CT \mathbb{CP}^{N-1})}{\hat\Gamma_c(\CN \CX)}\ ,
\end{equation}
so that
\begin{eqnarray}\nonumber
Z_{\mathbb{RP}^2}^{pert.} &=& C_0\int_\CX
e^{-iJ}~
\frac{\hat\Gamma_c\left(\frac{\CT \CX}{2}\right)
\hat\Gamma_c\left(-\frac{\CT \CX}{2}\right)}{\hat\Gamma_c(-\CT \CX)}\\
&=& C_0\int_\CX
e^{-iJ}~
\frac{\CA\left(\frac{\CT \CX}{2}\right)}{\hat\Gamma_c(-\CT \CX)}\ ,
\end{eqnarray}
where $\CA$ is the $\hat A$ class.
This shows that in the large volume limit,
the conventional overlap amplitude between
RR-ground state and a crosscap state (see e.g., \cite{Brunner:2003zm})
are corrected by replacing
\begin{equation}
\sqrt{\CL(\CT \CX/4)}\quad\rightarrow\quad \frac{\CA\left({\CT \CX}/{2}\right)}{\hat\Gamma_c(-\CT \CX)}\ .
\end{equation}
In section 6, we will come back to this expression and
explore the consequences.

\subsection{Orientifolds with a Normal Bundle}

Lower dimensional Orientifold planes, from B-parity projection,
may wrap a holomorphically embedded surface $\CM$ in the ambient Calabi-Yau $\CX$,
if $\CX$ admits $Z_2$ discrete symmetries.
At the level of GLSM, this is achieved by combining the parity
projection with such a flavor symmetry, as we considered
in section \ref{flavor}.

For example, the simplest such Calabi-Yau has a
superpotential $P\cdot G_N=P\cdot \sum_{a=1}^N X_i^N$ which is
invariant under exchange of $X$'s among themselves.
Exchanging  a pair of chiral
fields $X^1\leftrightarrow X^2$ gives rise to a fixed
locus defined by $X^1+X^2=0$, a complex
co-dimension one hypersurface as well as a complex co-dimension $(N-2)$ subspace,
i.e., a point at $X^3=\cdots=X^N=0$.
We can do the similar analysis for the symmetry
exchanging $X^1\leftrightarrow X^2$ and $X^3\leftrightarrow X^4$
simultaneously. This action gives complex co-dimension 2
fixed locus defined as $(X^1,\cdots X^N)=(X,X,Y,Y,X^5,\cdots, X^N)$,
and co-dimension $(N-3)$ fixed locus, $(X^1,\cdots X^N)=(X,-X,Y,-Y,0,\cdots, 0)$.
For the quintic, both of these correspond to $O5$ planes.
These results are summarized
in the following table \cite{Brunner:2003zm}.
\begin{equation}
\begin{array}{|c|c|}
\hline
(X_1,X_2,X_3,X_4 ,X_5) \rightarrow (X_1,X_2,X_3,X_4, X_5) & O9~(\mathrm{spacetime~filling})\\
\hline
(X_1,X_2,X_3,X_4 ,X_5) \rightarrow (X_2,X_1,X_3,X_4 ,X_5) & \begin{array}{c}O7~\mathrm{at}~(X,X,X_3,X_4,X_5)
\\ O3~\mathrm{at}~(X,-X,0,0,0) \end{array}\\
\hline
(X_1,X_2,X_3,X_4 ,X_5) \rightarrow (X_2,X_1,X_4,X_3, X_5) & \begin{array}{c}O5~\mathrm{at}~(X,X,Y,Y,X_5)
\\ O5~\mathrm{at}~(X,-X,Y,-Y,0)\end{array}\\
\hline
\end{array}
\end{equation}

As this shows, we generically end up with more than one
Orientifold planes, given a parity projection. The central
charges must be all present in the $\mathbb{RP}^2$ partition
function, so the latter must be  in general composed of
more than one additive terms. What allows this is the
holonomy sectors we encountered in section 3.
For a GLSM gauge group $U(n)$, for example, one
has $n+1$ such distinct holonomy sectors, and
can accommodate several Orientifold planes. For the current
example of $U(1)$ GLSM, we have exactly two such holonomy
sectors, and thus up to two Orientifolds planes.\footnote{
For the spacetime-filling case of section 5.1, only even sector
contributed to the large-volume limit, and there was
only one type of Orientifold plane. However, the odd
holonomy piece is still important in the
following sense: Thanks to the $U(1)$ gauge symmetry
of GLSM, one can alternatively project with $X\rightarrow -X$
and $P\rightarrow (-1)^N P$ without changing the theory.
However, this flips the even and the odd holonomy sector
precisely, which implies that the large-volume central
charge of the spacetime-filling Orientifold planes
resides in the odd holonomy sector instead. }

In the end, our examples below, combined
with the spacetime-filling case above, will suggest a universal formula
for the large volume central charge
\begin{eqnarray}\label{normalbundle}
Z_{\mathbb{RP}_2}^{pert.}&=&C_{-s}\int_\CX
e^{-iJ} ~\frac{\CA(\CT\CM/2)}{\hat\Gamma_c(-\CT\CM)}
\wedge
\frac{\hat\Gamma_c(\CN\CM)}{\CA(\CN\CM/2)}
\wedge e(\CN\CM)\cr
&=&C_{-s}\int_\CM
e^{-iJ} \wedge\frac{\CA(\CT\CM/2)}{\hat\Gamma_c(-\CT\CM)}
\wedge
\frac{\hat\Gamma_c(\CN\CM)}{\CA(\CN\CM/2)}
\ ,
\end{eqnarray}
for an Orientifold plane $\CM$  of real co-dimension $2s$ in a
Calabi-Yau $d$-fold $X$, with $C_{-s}= i^{d-s}2^{d-2s}~(2\pi)^{(d+2)/2}~(r\L)^{c/6}$.

\subsubsection{\bf Orientifold Planes of Complex Co-Dimensions $1$ \& $N-2$  }

Let us consider the projection involving $X^1\leftrightarrow X^2$.
As the table above shows, this produces two different
fixed planes; An hyperplane with $X^1=X^2$ and an isolated point
at $X^3=\cdots=X^N=0$. Thus, we expect to recover additive
contributions from these two planes,
for which existence of the two holonomy sectors is crucial.

As we are considering the ambient Calabi-Yau  $\CX$ as a
hypersurface embedded in $\mathbb{CP}^{N-1}$, the
results of section \ref{flavor} reads
\begin{equation}\label{ex1}
(2\pi)^{-N/2+2}~2^{-N}~(r\Lambda)^{c/6}~\mathrm{res}_{\e=0}
\frac{\Gamma(\e)}{
\Gamma\left(1-\e\right)}\cdot
\left[\frac{\Gamma\left(\frac{\epsilon}{2}\right)
\Gamma\left(-\frac{\epsilon}{2}\right)}{\Gamma(-\epsilon)}\right]^{N-2}
\cdot\left[\frac{\Gamma\left(\frac{1-N\epsilon}{2}\right)
\Gamma\left(-\frac{1-N\epsilon}{2}\right)}{\Gamma(-1+N\epsilon)}\right]\ ,
\end{equation}
from the even holonomy sector,
\begin{equation}\label{ex2}
(2\pi)^{N/2+1}~2^{N-2}~(r\Lambda)^{c/6}~\mathrm{res}_{\e=0}
\frac{\Gamma(\e)}{
\Gamma\left(1-\e\right)}\cdot
\left[\frac{\Gamma(\epsilon)/\epsilon}{\Gamma\left(\frac{\epsilon}{2}\right)
\Gamma\left(-\frac{\epsilon}{2}\right)}\right]^{N-2}\cdot
\left[\frac{\Gamma(1-N\epsilon)/(1-N\epsilon)}{\Gamma\left(\frac{1-N\epsilon}{2}\right)
\Gamma\left(-\frac{1-N\epsilon}{2}\right)}\right]\ ,
\end{equation}
from the odd holonomy sector. Note that, for this case,
both sectors contribute to the residue at $\e=0$.

First, let us consider the even holonomy sector contribution.
With (\ref{identity1}), we may write (\ref{ex1}) as
\begin{eqnarray}\nonumber
&&-(2\pi)^{-N/2+3}~2^{-N+2}~(r\Lambda)^{c/6}
~\mathrm{res}_{\e=0}~
\frac{1}{\e} \cdot \frac{\Gamma(\e)}{\Gamma\left(\frac{\e}{2}\right)
\Gamma\left(-\frac{\e}{2}\right)}\\
&&\phantom{aaaaaaaaaaaaaaaaaaaaaaa}
\times\left[\frac{\Gamma\left(\frac{\epsilon}{2}\right)
\Gamma\left(-\frac{\epsilon}{2}\right)}{\Gamma(-\epsilon)}\right]^{N-1}
\frac{\Gamma(-N\epsilon)}{\Gamma\left(\frac{N\epsilon}{2}\right)
\Gamma\left(-\frac{N\epsilon}{2}\right)}\\\nonumber
&=&(2\pi)^{-N/2+3}~2^{N-4}~(r\Lambda)^{c/6}~\mathrm{res}_{\e=0}~
 \frac{N}{\epsilon^{N-2}}\cdot
\frac{\Gamma(1+\e)}{\Gamma\left(1+\frac{\e}{2}\right)
\Gamma\left(1-\frac{\e}{2}\right)}\\
&&\phantom{aaaaaaaaaaaaaaaaaaaaaaa}\nonumber
\times
\left[\frac{\Gamma\left(1+\frac{\epsilon}{2}\right)
\Gamma\left(1-\frac{\epsilon}{2}\right)}{\Gamma(1-\epsilon)}\right]^{N-1}
\frac{\Gamma(1-N\epsilon)}{\Gamma\left(1+\frac{N\epsilon}{2}\right)
\Gamma\left(1-\frac{N\epsilon}{2}\right)}\ .
\end{eqnarray}
Expressing the residue integral at $\e=0$ via an integral over $CX$ with
the hyperplane class $H$, we find
\begin{eqnarray}
Z_{\mathbb{RP}_2}^{pert.,~even}&=&C_{-1} \int_\CX
e^{-i\xi H}\wedge H\wedge\left[
\frac{\Gamma(1+\frac{H}{2\pi i})}{\Gamma\left(1+\frac{H}{4\pi i}\right)
\Gamma\left(1-\frac{H}{4\pi i}\right)}\right]\\
&&\qquad\wedge
\left[\frac{\Gamma\left(1+\frac{H}{4\pi i}\right)
\Gamma\left(1-\frac{H}{4\pi i}\right)}{\Gamma(1-\frac{H}{2\pi i})}\right]^{N-1}
\Biggl/\left[\frac{\Gamma\left(1+\frac{NH}{4\pi i}\right)
\Gamma\left(1-\frac{NH}{4\pi i}\right)}{\Gamma(1-\frac{NH}{2\pi i})}\right]\ ,\nonumber
\end{eqnarray}
with  $C_{-1}= i^{N-3}(2\pi)^{N/2}2^{N-4}(r\Lambda)^{c/6}$.
Note that, again in terms of the $\CA$ and $\hat\Gamma_c$ classes,
this formula can be organized as
\begin{equation}\label{1}
Z_{\mathbb{RP}_2}^{pert., ~even}=C_{-1}\int_{\CM_{-1}}
e^{-iJ} \wedge\frac{\CA(\CT\CM_{-1}/2)}{
\hat\Gamma_c(-\CT\CM_{-1})}
\wedge
\frac{\hat\Gamma_c(\CN\CM_{-1})}{\CA(\CN\CM_{-1}/2)}
\ ,
\end{equation}
where $\CM_{-1}$ denotes for a complex co-dimension 1 fixed locus,
parameterized by $(X^1,\cdots X^N)=(X,X,X^3,\cdots X^N)$.

Contribution from the odd holonomy sector can be similarly written as
\begin{eqnarray}
&&(-1)^{N-1}2^{-N+2}~(2\pi)^{N/2}~(r\Lambda)^{c/6}~\mathrm{res}_{\e=0}
\frac{1}{\e}\cdot\frac{\Gamma\left(1+\frac{\e}{2}\right)
\Gamma\left(1-\frac{\e}{2}\right)}{\Gamma(1-\e)}\\ \cr
&&\phantom{aaaaaaaaaaaaaaaaaaaaaaa} \times \left[\frac{\Gamma(1+\epsilon)}{\Gamma\left(1+\frac{\epsilon}{2}\right)
\Gamma\left(1-\frac{\epsilon}{2}\right)}\right]^{N-1}\cdot
\frac{\Gamma\left(1+\frac{N\e}{2}\right)\Gamma\left(1-\frac{N\e}{2}\right)}{
\Gamma\left(1+N\e\right)}\ ,\nonumber
\end{eqnarray}
which is equivalent to
\begin{eqnarray}
Z_{\mathbb{RP}_2}^{pert.,~odd}&=&C_{-(N-2)}\int_\CX
e^{-iJ}\wedge\frac{ H^{N-2}}{N}\wedge
\frac{\Gamma\left(1+\frac{H}{4\pi i}\right)
\Gamma\left(1-\frac{H}{4\pi i}\right)}{\Gamma(1-\frac{H}{2\pi i})}\\
&\wedge&\left[\frac{\Gamma(1+\frac{H}{2\pi i})}{\Gamma\left(1+\frac{H}{4\pi i}\right)
\Gamma\left(1-\frac{H}{4\pi i}\right)}\right]^{N-1}
\Biggl/\left[\frac{\Gamma\left(1+\frac{NH}{2\pi i}\right)}{
\Gamma\left(1+\frac{NH}{4\pi i}\right)\Gamma\left(1-\frac{NH}{4\pi i}\right)
}\right]\ , \nonumber
\end{eqnarray}
where $C_{-(N-2)}=(-1)^{N-1}2^{-N+2}~(2\pi)^{N/2}~(r\L)^{c/6}$.
Again we may rewrite this as an integral
\begin{equation}\label{N-2}
Z_{\mathbb{RP}_2}^{pert., ~odd}=C_{-(N-2)}\int_{\CM_{-(N-2)}}
e^{-iJ} ~\frac{\CA(\CT\CM_{-(N-2)}/2)}{\hat\Gamma_c(-\CT\CM_{-(N-2)})}
\wedge
\frac{\hat\Gamma_c(\CN\CM_{-(N-2)})}{\CA(\CN\CM_{-(N-2)}/2)}
\ ,
\end{equation}
over $\CM_{N-2}$ which, in this case, is actually evaluation at
the fixed point at $(X^1,\cdots X^N)=(X,-X,0,\cdots,0)$.

\subsubsection{\bf Orientifold Planes of  Complex Co-Dimensions $2$ \& $N-3$  }


Next, we consider the B-parity action that exchanges
$X^{1}\leftrightarrow X^2$ and $X^{3} \leftrightarrow X^{4}$
simultaneously. Similarly, from the even holonomy sector,
we have
\begin{equation}
(2\pi)^{-N/2+3}~2^{-N+2}~(r\Lambda)^{c/6}~\mathrm{res}_{\e=0}
\left[\frac{\Gamma(\e)}{
\Gamma\left(1-\e\right)}\right]^2
\left[\frac{\Gamma\left(\frac{\epsilon}{2}\right)
\Gamma\left(-\frac{\epsilon}{2}\right)}{\Gamma(-\epsilon)}\right]^{N-4}
\left[\frac{\Gamma\left(\frac{1-N\epsilon}{2}\right)
\Gamma\left(-\frac{1-N\epsilon}{2}\right)}{\Gamma(-1+N\epsilon)}\right]\ ,
\end{equation}
and from the odd holonomy sector,
\begin{equation}
(2\pi)^{N/2}~2^{N-4}~(r\Lambda)^{c/6}~\mathrm{res}_{\e=0}
\left[\frac{\Gamma(\e)}{
\Gamma\left(1-\e\right)}\right]^2
\left[\frac{\Gamma(\epsilon)/\epsilon}{\Gamma\left(\frac{\epsilon}{2}\right)
\Gamma\left(-\frac{\epsilon}{2}\right)}\right]^{N-4}
\left[\frac{\Gamma(1-N\epsilon)/(1-N\epsilon)}{\Gamma\left(\frac{1-N\epsilon}{2}\right)
\Gamma\left(-\frac{1-N\epsilon}{2}\right)}\right]\ .
\end{equation}
Again, both  holonomy sectors contribute for the residue at
$\epsilon=0$.

For the even holonomy sector, a similar procedure
gives
\begin{eqnarray}
Z_{\mathbb{RP}_2}^{pert.,~even}&=&C_{-2} \int_\CX
e^{-i\xi H}\wedge H^2\wedge\left[
\frac{\Gamma(1+\frac{H}{2\pi i})}{\Gamma\left(1+\frac{H}{4\pi i}\right)
\Gamma\left(1-\frac{H}{4\pi i}\right)}\right]^2\\
&&\qquad\wedge
\left[\frac{\Gamma\left(1+\frac{H}{4\pi i}\right)
\Gamma\left(1-\frac{H}{4\pi i}\right)}{\Gamma(1-\frac{H}{2\pi i})}\right]^{N-2}
\Biggl/\left[\frac{\Gamma\left(1+\frac{NH}{4\pi i}\right)
\Gamma\left(1-\frac{NH}{4\pi i}\right)}{\Gamma(1-\frac{NH}{2\pi i})}\right]\ .\nonumber
\end{eqnarray}
where $C_{-2}=i^{N-4}(2\pi)^{N/2}2^{N-6}(r\Lambda)^{c/6}$.
In terms of the characteristic classes, we rewrite this
\begin{equation}\label{2}
Z_{\mathbb{RP}_2}^{pert.,~even}=C_{-2}\int_{\CM_{-2}}
e^{-iJ} \wedge\frac{\CA(\CT\CM_{-2}/2)}{\hat\Gamma_c(-\CT\CM_{-2})}
\wedge
\frac{\hat\Gamma_c(\CN\CM_{-2})}{\CA(\CN\CM_{-2}/2)}
\ ,
\end{equation}
with $\CM_{-2}$ is complex co-dimension 2 fixed locus,
$(X_1,\cdots X_N)=(X,X,Y,Y,X_5,\cdots X_N)$.

Finally, from the odd holonomy sector, we have
\begin{eqnarray}\nonumber
Z_{\mathbb{RP}_2}^{pert.,~odd}&=&C_{-(N-3)}\int_\CX
e^{-i\xi H}\wedge\frac{ H^{N-3}}{N}\wedge
\left[\frac{\Gamma\left(1+\frac{H}{4\pi i}\right)
\Gamma\left(1-\frac{H}{4\pi i}\right)}{\Gamma(1-\frac{H}{2\pi i})}\right]^2\\
&\wedge&\left[\frac{\Gamma(1+\frac{H}{2\pi i})}{\Gamma\left(1+\frac{H}{4\pi i}\right)
\Gamma\left(1-\frac{H}{4\pi i}\right)}\right]^{N-2}
\Biggl/\left[\frac{\Gamma\left(1+\frac{NH}{2\pi i}\right)}{
\Gamma\left(1+\frac{NH}{4\pi i}\right)\Gamma\left(1-\frac{NH}{4\pi i}\right)
}\right]\ ,
\end{eqnarray}
where $C_{-(N-3)}=i(-1)^{N-1}(2\pi)^{N/2}~2^{-N+4}$. This again can be
summarized as
\begin{equation}\label{N-3}
Z_{\mathbb{RP}_2}^{pert.,~odd}=C_{-(N-3)}\int_{\CM_{-(N-3)}}
e^{-iJ}\wedge \frac{\CA(\CT\CM_{-(N-3)}/2)}{\hat\Gamma_c(-\CT\CM_{-(N-3)})}
\wedge
\frac{\hat\Gamma_c(\CN\CM_{-(N-3)})}{\CA(\CN\CM_{-(N-3)}/2)}
\ ,
\end{equation}
where $\CM_{-(N-3)}$ is a co-dimension $N-3$ locus spanned by
$(X,-X,Y,-Y,0,\cdots,0)$.

\section{Consistency Checks and Subtleties}

In this last section, we explore the disk amplitudes
${}_R\langle 0| \CB\rangle_R$ and the crosscap amplitudes
${}_R\langle 0| \CC\rangle_R$ further.
The most immediate question is whether these
two types of amplitudes, or equivalently the central charges,
come out with the correct relative normalization, for which we
kept the overall coefficients carefully in the above. We will
then ask subtler questions of what should happen when $\CM$
is not Spin but only Spin$^c$, for which we can only offer
a guess for the final expression but not a derivation.

We then move on to the anomaly inflow and also  how we should
extract, from the computed central charge,  the RR-tensor
Chern-Simons coupling. Having both  ${}_R\langle 0| \CB\rangle_R$
and  ${}_R\langle 0| \CC\rangle_R$ explicitly is most telling
in this regard, whereby we discover that the difference between
the conventional central charges and the newly computed ones is
universal; the extra multiplicative factor due to $\hat\Gamma_c$
class is common for both D-branes and Orientifold planes and the
same again makes appearance in $S^2$ partition function as well.
This strongly suggests that the change should be attributed to
the quantum volume of the cycles in $\CX$, rather than to the
characteristic class that appears in the world-volume Chern-Simons
coupling to the spacetime RR tensor fields.

\subsection{Tadpole }\label{Tadpole}

The simplest consistency check comes from the tadpole cancelation condition of
the RR ground states, which can be written as \cite{Polchinski:1987tu,Brunner:2004zd}
\begin{equation}\label{tcondition}
{}_R\langle 0| \CC\rangle_R +{}_R\langle 0| \CB\rangle_R =0\ ,
\end{equation}
and demand the boundary state be constrained to satisfy this equality.
{}From the spacetime viewpoint, this is the Gauss constraint for
the RR-tensor fields, integrated over the compact Calabi-Yau manifold.  Recall that
the RR-charge of a single D$p$-brane and that of an $Op^\pm$ Orientifold plane must have a
relative weight of
\begin{equation}
\pm 2^{p-4}
\end{equation}
in the covering space. Obviously, the same numerical factor must appear in
the central charges.

For this numerical factor, we start with Hori and Romo \cite{Hori:2013ika},
and consider tachyon condensation to obtain the disk partition function for
a D-brane wrapping $\CM$ in $\CX$
\begin{equation}\label{disk}
Z_{D^2}=
(r\Lambda)^{c/6}~(2\pi)^{(d+2)/2}\int_\CM e^{-B-iJ}\wedge ch(\CE)
\wedge\frac{\hat\Gamma_c(\CT)}{\hat\Gamma_c(-\CN)}\wedge e^{-c_1(\CN)/2}\ ,
\end{equation}
where $d$ is the complex dimension of the Calabi-Yau $\CX$. See Appendix C
for details of this procedure. On the other hand, the result of section \ref{CY}
can be written as
\begin{equation}\label{normalbundle2}
Z_{\mathbb{RP}^2}
= ~2^{d-2s}(r\Lambda)^{c/6}(2\pi)^{(d+2)/2}
\int_\CM e^{-iJ}\wedge
\frac{\CA(\CT/2)}{\hat\Gamma_c(-\CT)}
\wedge
\frac{\hat\Gamma_c(\CN)}{\CA(\CN/2)}
\ ,
\end{equation}
where the complex co-dimension of $\CM$ is denoted by $s$.
The last factor in \eqref{disk} and its apparent absence
in \eqref{normalbundle2} is the subject of the next
subsection; for tadpole issue, it suffices to know that
the 0-form part of the two expressions differ by the
numerical factor of rank($\CE$), prior to the projection,
and also by $2^{d-2s}$. For the familiar Ramond-Ramond
tadpole cancelation condition to emerge correctly,
therefore, $2^{d-2s}$ must equal $2^{p-4}$.
For ten-dimensional spacetime, $d=10/2=5$ and $p=9-2s$,
so $d-2s=p-4$, precisely as needed.

\subsection{A Subtlety with Spin$^c$ Structure}

A well-known subtlety with D-branes occurs when they wrap a
manifold $\CM$ which is not Spin. This causes a global anomaly in 2D
boundary CFT, whereby the world-sheet fermion determinant has
an ill-defined sign. As pointed out by Freed and Witten \cite{Freed:1999vc}
this ambiguity is cancelable by additional phase factor,
provided that $\CM$ is Spin$^c$,
\begin{equation}
\exp\left ({i\int_{\partial\Sigma} \hat A}\right) \ ,
\end{equation}
with some world-volume Abelian ``gauge field" $\hat A$. The latter
is equally ill-defined, precisely such that the sign flip due to
the world-sheet global anomaly is canceled by the sign ambiguity
of the latter.

A related observation is that
spacetime spinor is ill-defined on a Spin$^c$ manifold, which
is nevertheless correctable if we think of the spinor as a section
of $\hat L^{1/2}\,\otimes\, \CS(\CT\CM)$, where  $\hat A$ is the ``connection"
on the ill-defined bundle $\hat L^{1/2}$. This implies that the
Dirac index on $\CM$ is equally ill-defined unless we twist
the Dirac operator by $\hat L^{1/2}$ and once this
is done we have an index theorem,
\begin{equation}
\int_{\CM} e^{\hat F/2\pi}\wedge\CA(\CT\CM)\wedge \cdots
\end{equation}
with $\hat  F=d \hat A$, where the ellipsis denotes contributions
from the well-defined part of the gauge bundle.
A little experiment with this index formula\footnote{With the
aim at obtaining integer values of the index for completely
smooth an compact examples like $\mathbb{CP}^{2k}$ or other
toric Spin$^c$ manifold. See also \cite{Katz:2002gh}.} suggests that a good de Rham cohomology
representative  for $\hat F/2\pi$ is $c_1(\CM)/2$. One can
understand this from the fact that it is $c_1(\CM)$,
or more precisely the 2nd Stiefel-Whitney class
$$w_2(\CM)\quad= \quad c_1(\CM)\;\; \hbox{mod}\;\; Z_2$$
that
determines whether the manifold is Spin. With $w_3(\CM)=0$,
therefore, $c_1(\CM)/2$ determines whether the manifold is
Spin or Spin$^c$.

For $\CM$ embedded in an Calabi-Yau ambient $\CX$ so that
$c_1(\CT)+c_1(\CN)=0$, this implies an additional factor
\begin{equation}
e^{\hat F/2\pi}=e^{-c_1(\CN)/2}
\end{equation}
in the central charge (and in the RR-charge) of the D-brane,
whose presence was argued by Minasian and Moore \cite{Minasian:1997mm}: the
correct central charge  must have this extra factor,
\begin{equation}\label{disk2}
Z_{D^2}\sim \int_\CM e^{-B-iJ}\wedge ch(\CE)\wedge
\cdots
\wedge e^{-c_1(\CN)/2}\ .
\end{equation}
In view of its origin as the ``half line
bundle" $\hat L^{1/2}$, it makes more sense to think of it as
part of the ``gauge bundle" $\CE\rightarrow \CE\otimes
\hat L^{1/2}$.

When $\CM$ is Spin, however, this is a mere redefinition of
$\CE$ since $\hat L^{1/2}$ is a proper line bundle
when $\hat F/2\pi = c_1(\CM)/2$ is integral. The D-brane
spectra is, as expected, not affected by such factor
when $\CM$ is Spin. For this reason (and also because
the Orientifold cannot admit gauge bundles), the right thing
to do is to keep this factor explicitly only when
$\CM$ is Spin$^c$. With this in mind, we will write, instead
\begin{equation}\label{disk3}
Z_{D^2}\sim \int_\CM e^{-B-iJ}\wedge ch(\CE)\wedge
\cdots
\wedge e^{d(\CM)/2}\ ,
\end{equation}
where
$$
d(\CM)=\left\{\begin{array}{ll} 0\quad &\CM \;\hbox{ is Spin} \\
c_1(\CT\CM)=-c_1(\CN\CM)\quad &\CM\; \hbox{ is Spin}^c
\end{array}\right\}\ ,
$$
again by redefinition of the gauge bundle $\CE$.

For D-branes, Appendix C outlines how one can compute
the hemisphere partition functions, starting
with the result in \cite{Hori:2013ika,Honda:2013uca}, via the tachyon
condensation. In this approach, one does find the factor
$e^{-c_1(\CN)/2}$, where the key point lies with charge
assignment for the Hilbert space vacua
\cite{Herbst:2008jq,Hori:2013ika,Honda:2013uca}
associated with the boundary degrees of freedom.
With ``correct" choice of the charges, we find Eq.~\eqref{disk}.
In view of the global anomaly, this result is quite
natural. Since the original Calabi-Yau manifold is always Spin
and thus free of the global anomaly,
the lower dimensional D-brane induced from it must be equipped
with the necessary twist to countermand the potential  anomaly
on the induced D-brane, as it must flow to a well-defined
boundary CFT again.

However, if one imposes the Dirichlet boundary condition
from the outset, to obtain  lower dimensional D-branes
in the hemisphere partition function \cite{Hori:2013ika},
the origin of such a factor is at best subtle. The naive computation
from imposing the Dirichlet boundary condition, in contrast to the
tachyon condensation above, does not seem to generate
the factor in question. Again, we refer the readers
to Appendix C for discussion on the matter.

The global anomaly and the resulting subtlety with
Spin$^c$ manifold must also exist for Orientifold planes.
Distler, Freed, and Moore \cite{Distler:2009ri,Distler:2010an}
have stated that a similar global anomaly is present but canceled
by the sign ambiguity of the factor
\begin{equation}
\exp\left ({i\int_{\Sigma} B}\right) \ ,
\end{equation}
although, because one works with non-orientable manifolds,
even the definition of this expression requires more work.
An important evidence that favors the same extra factor
on Orientifolds is the anomaly inflow onto D-branes
and I-branes. See next subsection for how such a coupling on
D-brane world-volume basically demands the same factor to
appear on the Orientifold world-volume.

Its origin for the Orientifolds is however even less
clear than the D-brane case. For one thing, the tachyon condensation does not yield
Orientifold planes. There must be subtleties with $B$ field coupling
that should be responsible for this, which we are yet
to understand properly. In the next subsection,
we will see how simultaneous restoration of this factor on both
D-branes and Orientifold planes is consistent with
anomaly inflow need to cancel world-volume anomaly. For large
volume central charges with Calabi-Yau $\CX$ , this involves
multiplying a factor $e^{d(\CM)/2}$ on the right
hand sides of Eqs. \eqref{normalbundle}, \eqref{1}, \eqref{N-2},
\eqref{2},  \eqref{N-3}, and \eqref{normalbundle2}.

\subsection{Anomaly Inflow and Indices}\label{inflow}

Let $|a\rangle_{RR}$ denote one of the crosscap or boundary states
in the Ramond-Ramond sector. Then one can naturally define the Witten index as
\begin{equation}
I(a,b) =\lim_{T\rightarrow \infty}~{}_{RR}\langle a|e^{-TH}|b\rangle_{RR}\ ,
\end{equation}
which calculates the indices of open strings attached between
D-branes and Orientifold planes. Following figures are three
distinguished topologies which give rise to the indices for
brane-brane, brane-plane, and plane-plane
respectively.

Due to the Riemann bilinear identity, these indices can be
expressed in terms of the partition functions as follows \cite{Brunner:2003zm}.

\begin{eqnarray}\label{bb}
I(\CB_E,\CB_F) &=& \sum_{ij}\langle \CB_E|i\rangle\eta^{ij}\langle j |\CB_F\rangle\ ,\\\label{bc}
I(\CB_E,\CC)&=& \sum_{ij}\langle \CB_E|i\rangle\eta^{ij}\langle j |\CC\rangle\ ,\\\label{cc}
I(\CC,\CC)&=& \sum_{ij}\langle \CC|i\rangle\eta^{ij}\langle j |\CC\rangle\ ,
\end{eqnarray}
where all the states are in the Ramond-Ramond sector,
and $\eta^{ij}$ is the topological metric of the chiral ring elements.
Since the overlap between the RR ground states and the boundary/crosscap
states measures the coupling to the RR gauge fields, this formula
can be thought of as inflow mechanism which cancels the one-loop
anomaly from each open string sector. Since the expression for these
indices in the geometric limit are well-known in the literature, we can
check whether our results generate expected indices, and consistency
with the original inflow mechanism
\cite{GHM,CY,Minasian:1997mm}.

\begin{figure}\label{fig}
\centering
\includegraphics[width=0.3\textwidth]{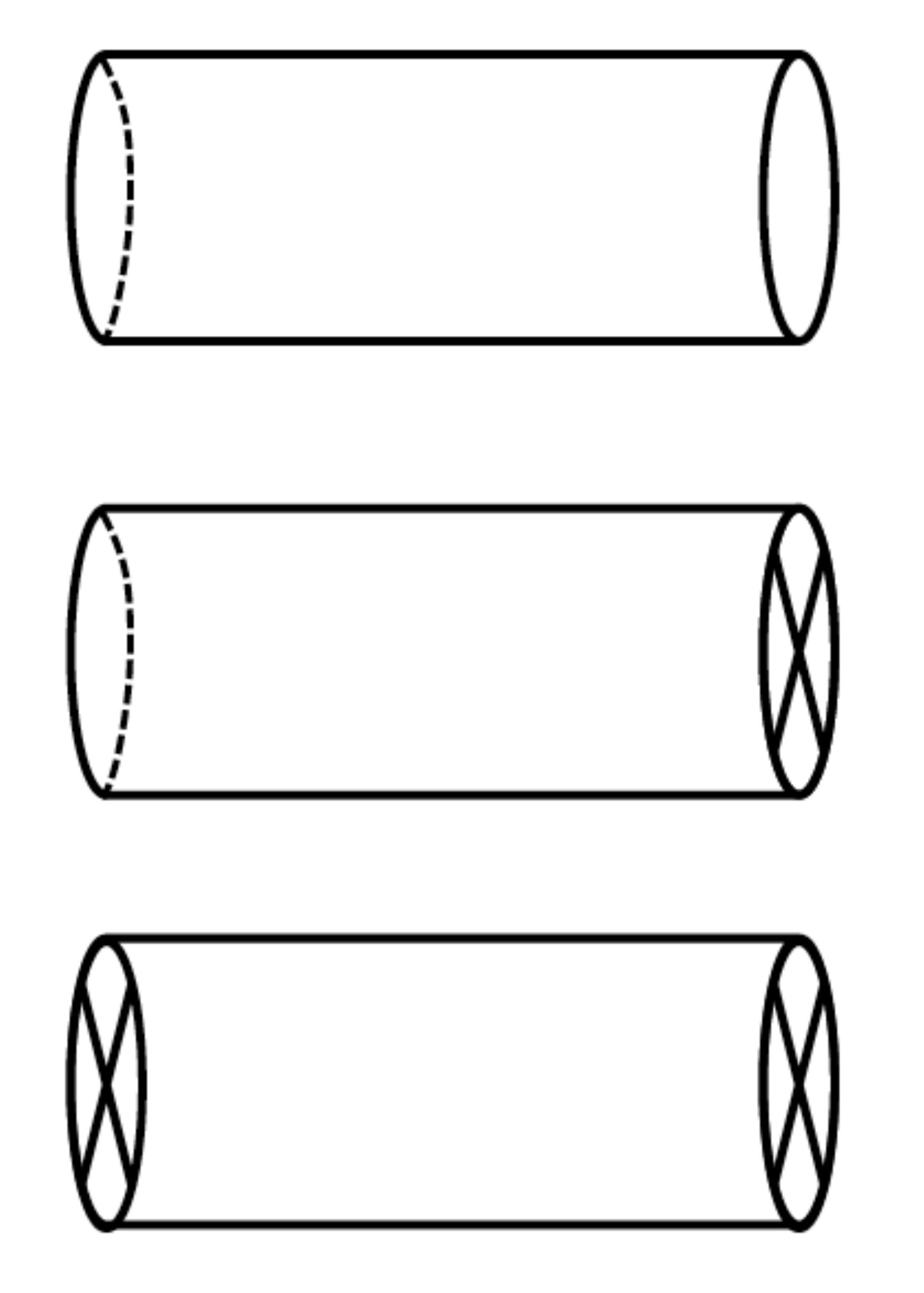}
\caption{Two dimensional topologies where indices are defined.
The first one denotes for a cylinder with two boundaries at the ends,
and the second one corresponds to the M\"obius strip with one boundary
and one crosscap. The last one is the Klein bottle, with two crosscap
states at the ends. }
\end{figure}

Following the discussion of the previous subsection,
here we assume that an extra factor
$e^{d(\CM)/2}$ is present not only on the world-volumes of D-branes
but also on the world-volumes of Orientifold planes.
Otherwise, amplitudes
involving boundary states only and amplitude involving a
boundary state and a crosscap cannot be summed up; this  would
lead to net world-volume anomaly and make the spacetime theory
inconsistent.
Because we assume $\CX$ itself to be Spin, $d(\CM)/2$ is always
expressed as a sum over $-c_1/2$ of the normal bundles of the world-volumes.
As we have not demonstrated the GLSM origin of this factor for Orientifolds,
the readers may wish to regard the following with the assumption
of $d=0$, that is, only for Spin $\CM$'s.

\paragraph{Cylinder}

Index on the cylinder and relation to the disk partition function were
studied in \cite{Hori:2013ika} and \cite{Honda:2013uca}.

We start with Eq.~(\ref{disk})
and use the relation (\ref{bb}) to calculate the open string
index stretched between two branes with $(\CE_1,\CM_1)$
and $(\CE_2,\CM_2)$ as
\begin{eqnarray}\nonumber
&&I(\CB_{\CE_1},\CB_{\CE_2}) \\\nonumber
&\sim &\int_{\CM_1\cap\CM_2} e^{-B-iJ}\wedge
ch(\CE_1)\wedge\frac{\hat\Gamma_c(\CT_1)}{\hat\Gamma_c(-\CN_1)}
\wedge e^{d(\CM_1)/2}\\\label{DDanomaly}
&&\phantom{qqqqqqq}
\wedge e^{B+iJ}
\wedge ch(-\CE_2)\wedge\frac{\hat\Gamma_c(-\CT_2)}{\hat\Gamma_c(\CN_2)}
\wedge e^{-d(\CM_2)/2}
\wedge e(\CN_{12})\\\nonumber
&=&\int_{\CM_1\cap\CM_2}
ch(\CE_1)\wedge ch(-\CE_2)\wedge \frac{\CA(\CT(\CM_1\cap\CM_2))}{\CA(\CN(\CM_1\cap\CM_2))}
\wedge e^{\left(d(\CM_1)-d(\CM_2)\right)/2}
\wedge e(\CN_{12})\ ,
\end{eqnarray}
where $\CT_i$ and $\CN_i$ denote for tangent and normal bundles
of $\CM_i$ and $\CN_{12}\equiv\CN_1\cap \CN_2$. From the first to
the second line, we used
\begin{eqnarray}
\frac{\hat\Gamma_c(\CT_1)\wedge\hat\Gamma_c(-\CT_2)}{\hat\Gamma_c(-\CN_1)\wedge\hat\Gamma_c(\CN_2)}
=
\frac{\hat\Gamma_c(\CT_1\cap\CT_2)\hat\Gamma_c(-\CT_1\cap\CT_2)}{
\hat\Gamma_c(-\CN_1\cap\CN_2)\hat\Gamma_c(\CN_1\cap\CN_2)}
=
\frac{\CA(\CT_1\cap\CT_2)}{\CA(\CN_1\cap\CN_2)}\ ,
\end{eqnarray}
since
\begin{equation}
\CT_1\backslash(\CT_1\cap\CT_2) = \CN_2\backslash(\CN_1\cap\CN_2)\ .
\end{equation}
Note that, for the first equality, complex conjugation of the
normal bundle in the denominator of Eq.~(\ref{disk}) is essential.

The factor $e^{\left(d(\CM_1)-d(\CM_2)\right)/2}$ in (\ref{DDanomaly}) can
be understood from the fact that the I-brane fermions on $\CM_1\cap\CM_2$ are naturally
sections of $\CS(\CT_1\cap \CT_2 \oplus \CN_1\cap \CN_2)$. When the latter
fails to be Spin, the 2nd Stiefel-Whitney class that measures this
failure is
$$
 w_2(\CT_1\cap \CT_2 \oplus
\CN_1\cap \CN_2)= w_2(\CT_1)-w_2(\CT_2) \ ,
$$
where the equality follows from the assumption that the ambient $\CX$
is Spin. Since $w_2=c_1$ mod $Z_2$, the relevant correcting
factor for the Spin$^c$ case is $e^{(c_1(\CT_1)-c_1(\CT_2))/2}$.
Note that this factor reduces to 1 when $\CM_1$ and $\CM_2$ are coincident,
which is expected since $\CT\oplus\CN=\CT\CX$ is Spin. Next, we show how
this extends to amplitudes involving Orientifold planes.

\paragraph{M\"obius strip}

Similarly, the index on the M\"obius strip can be obtained via
the relation (\ref{bc}). If we let $\CM_1$ and $\CM_2$ are locus
where D-branes and Orientifolds exist, we have\footnote{
From the first to second line, we used the identity
$
\sqrt{{\cal A}({\cal T})}\sqrt{{\cal L}({\cal T}/4)} ={\cal A}({\cal T}/2)\  .
$}
\begin{eqnarray}\nonumber
I(\CB_\CE,\CC)
&\sim &2^{p-4}\int_{\CM_1\cap\CM_2} 
ch_{2k}({\cal F})
\wedge \frac{\hat\Gamma({\CT_1})}{\hat\Gamma(-{\CN_1})}
\wedge \frac{{\cal A}({\CT_2}/2)\hat\Gamma(-{\CN_2})}{{\cal A}({\CN_2}/2)\hat\Gamma({\CT_2})}
\wedge e^{\left(d(\CM_1)-d(\CM_2)\right)/2}
\wedge e(\CN_{12})\\
&=&2^{p-4}\int_{\CM_1\cap\CM_2} 
ch_{2k}({\cal F}) \wedge
\sqrt{\frac{\CA(\CT_1)\CL(\CT_2/4)}{\CA(\CN_1)\CL(\CN_2/4)}}
\wedge e^{\left(d(\CM_1)-d(\CM_2)\right)/2}
\wedge e({\cal N}_{12})\ ,
\end{eqnarray}
which exactly reproduce the index formula of the M\"obius strip
calculated at the level of non-linear sigma model \cite{hep-th/9812071,Brunner:2003zm}. Here, $p+1$ is the
dimension of the Orientifold plane.

When D$p$-branes are on the top of an O$p$-plane, in particular, we can
read off $p+3$-form from $I(\CB_E,\CC) + I(\CC, \CB_E)$,
which gives anomaly inflow on the $p+1$ dimensional world-volume as
\begin{eqnarray}\nonumber
&&\pm~ 2^{p-4} \cdot  [ch_{2k}({\cal F})+ch_{\overline{2k}}({\cal F})]
\wedge \frac{\hat\Gamma({\cal T})}{\hat\Gamma(-{\cal N})}
\wedge \frac{{\cal A}({\cal T}/2)\hat\Gamma(-{\cal N})}{{\cal A}({\cal N}/2)\hat\Gamma({\cal T})}
\wedge e(\CN)~\Bigg|_{p+3}
\\\nonumber
&=&\pm ~ 2^{p-4} \cdot\left[
ch_{2k}({\cal F})+ch_{\overline{2k}}({\cal F})\right] \wedge
\frac{{\cal A}({\cal T}/2)}{{\cal A}({\cal N}/2)}\wedge
e({\cal N})~\Bigg|_{p+3}\\\label{bc+cb}
&=&\pm ~ ch_{2k}(2{\cal F})\wedge
\frac{{\cal A}({\cal T})}{{\cal A}({\cal N})}\wedge
e({\cal N})~\Bigg|_{p+3}\ .
\end{eqnarray}
Note that, since $U(k)$ gauge group is enhanced to $SO(2k)$
or $Sp(k)$ group, we used the relation $ch_{2k}(\CF)=ch_{\overline{2k}}(\CF)$.
Adding two contributions from the cylinder and the M\"obius indices,
we recover the open string Witten index, i.e., anomaly inflow
for the $SO(2N)$ or $Sp(N)$ gauge group according to the
sign of (\ref{bc+cb}),
\begin{eqnarray}\nonumber
I_{SO(2k),Sp(k)}&=&[ch_{2k\otimes 2k}(\CF)\pm ch_{2k}(2\CF)]\wedge \frac{\CA(\CT)}{\CA(\CN)}\wedge
e(\CN)\Bigg|_{p+3}\\
&=&2\cdot ch_{\frac12 2k(2k\pm1)}\wedge \frac{\CA(\CT)}{\CA(\CN)}\wedge
e(\CN)\Bigg|_{p+3}\ .
\end{eqnarray}

\paragraph{Klein bottle}
Finally, if there are two crosscap states as in the last diagram of
the figure, we have topology of the Klein bottle whose index is given
by the relation (\ref{cc}). Substituting our formula for the crosscap
overlap into this identity, we have
\begin{eqnarray}\nonumber
I(\CC,\CC) &\sim & 2^{p_1+p_2-8} \int_{\CM_1\cap\CM_2}
\frac{{\cal A}({\cal T}_1/2)\hat\Gamma({\cal N}_1)}{{\cal A}
({\cal N}_1/2)\hat\Gamma(-{\cal T}_1)}\wedge
\frac{{\cal A}({\cal T}_2/2)\hat\Gamma(-{\cal N}_2)}{{\cal A}({\cal N}_2/2)
\hat\Gamma({\cal T}_2)}
\wedge e^{\left(d(\CM_1)-d(\CN_2)\right)/2}
\wedge e({\cal N}_{12})\\\nonumber
&=& 2^{p_1+p_2-8}\int_{\CM_1\cap\CM_2}
\left(\frac{{\cal A}({\cal T}_1\cap\CT_2/2)}{{\cal A}({\cal N}_1\cap\CN_2/2)}
\right)^2
\wedge \frac{{\cal A}({\cal N}_1\cap\CN_2)}{{\cal A}({\cal T}_1\cap\CT_2)}
\wedge e^{\left(d(\CM_1)-d(\CM_2)\right)/2}
\wedge e({\cal N}_{12})\\
&=& 2^{p_1+p_2-8} \int_{\CM_1\cap\CM_2}
\frac{{\cal L}(\CT_1\cap\CT_2/4)}{{\cal L}(\CN_1\cap\CN_2/4)}
\wedge e^{\left(d(\CM_1)-d(\CM_2)\right)/2}
\wedge
e({\cal N}_{12})\ .
\end{eqnarray}
This again gives the well-known formula for the Klein bottle
index calculated in non-linear sigma model. Since the B-type
parity action corresponds to the Hodge star operation of the
target space, it reproduces the Hirzebruch signature theorem \cite{Brunner:2003zm}.
Obviously, this index is independent of the open string degrees
of freedom, or the types of planes \cite{Kim:2012wc}.
For type-I string theory, this inflow precisely cancels
the one-loop anomaly of supergravity multiplet.

\subsection{RR-Charges and Quantum Volumes}

This brings us, finally, to a natural question of what part
of the central charge should be attributed to the RR-charges.
Recall that the
conventional RR-charges, or the Chern-Simons coupling to
RR-tensors, was deduced indirectly via anomaly inflow.
For instance, for the simplest case of the spacetime-filling
D-brane, the relevant anomaly polynomial is $\CA(\CT)$, the
$\hat A$ class, which is then reconstructed via inflow
as
\begin{equation}\label{OOB}
\Omega(\CT)\wedge \Omega(-\CT)= \CA(\CT)\ ,
\end{equation}
where $\Omega$ is the characteristic class that appears in the
Chern-Simons coupling. With an implicit assumption that $\log\Omega $
is ``even," i.e., includes $4k$-forms only, this leads to $\Omega=\CA^{1/2}$
\cite{GHM,CY,Minasian:1997mm}. Some of early literatures were casual
about distinction between $\Omega(\CT)$ and $\Omega(-\CT)$,
although more careful computations show the conjugation
has to occur for one of the two factors \cite{CY,Kim:2012wc}.
Thus, in hindsight, the anomaly cancelation argument fixes
only ``even" part of $\log\Omega$.

As was noted previously, $\Omega=\hat\Gamma_c$ is one multiplicative
class that is consistent with the anomaly inflow $\CA$
in the above sense. This happens precisely because ``even"
part of $\log\hat\Gamma_c$ coincides exactly with $\log \CA^{1/2}$.
Our discussion in the previous section demonstrated that replacements like
\begin{equation}
\CA^{1/2}(\CT) \rightarrow
\hat\Gamma_c(\CT),\qquad \CL^{1/2}(\CT/4)
\rightarrow\CA(\CT/2)/\hat\Gamma_c(-\CT)\ ,
\end{equation}
for D-branes and Orientifold planes, respectively, would be
still consistent with anomaly inflow. However, since the central
charge is made from RR-charges and quantum volumes of various cycles,
it is hardly clear whether such a change in the central charge
should be attributed to the RR-charge or not.

More generally, for a D-brane wrapping a cycle $\CM$
in Calabi-Yau $\CX$, the gravitational curvature contribution
to the central charge is
\begin{equation}
\frac{\hat \Gamma_c(\CT)}{\hat \Gamma_c(-\CN)}=\sqrt{\frac{\CA(\CT)}{\CA(\CN)}}\wedge
\exp\left(\,i\, \sum_{k\geq 1}\frac{(-1)^k(2k)!\zeta(2k+1)}{(2\pi)^{2k}}ch_{2k+1}(\CX)\right) \ ,
\end{equation}
so the deviation depends only on $\CX$. As shown in the present work, something quite similar
happens for the Orientifold planes,
\begin{equation}
\frac{\CA(\CT/2)}{\hat \Gamma_c(-\CT)}\frac{\hat \Gamma_c(\CN)}{\CA(\CN/2)}=\sqrt{\frac{\CL(\CT/4)}{\CL(\CN/4)}}\wedge
\exp\left(\,i\, \sum_{k\geq 1}\frac{(-1)^k(2k)!\zeta(2k+1)}{(2\pi)^{2k}}ch_{2k+1}(\CX)\right) \ ,
\end{equation}
where the deviation is identical to its D-brane counterpart.
So the difference between the new central charges and the
conventional ones can be expressed by a universal factor,
determined by $\CX$ only, is independent of the choice of
the cycle $\CM$, and its logarithm is purely imaginary.

These properties all suggest that this factor should
be interpreted as a $\alpha'$ modification of volumes,
in the sense,
\begin{equation}\label{QK}
\exp(-iJ)\quad\rightarrow\quad \exp\left(-iJ+\,i\, \sum_{k\geq 1}
\frac{(-1)^k(2k)!\zeta(2k+1)}{(2\pi)^{2k}}ch_{2k+1}(\CX)\right)\ ,
\end{equation}
rather than as a shift of RR-charges, or the Chern-Simons couplings, themselves.
In fact, this is precisely the same shift of $J$ that appears in $S^2$
partition function, or its large volume expression,
\begin{eqnarray}
Z_{S^2}&\sim& \int_\CX e^{-2iJ}\wedge  \frac{\hat \Gamma_c(\CT\CX)}{\hat\Gamma_c(-\CT\CX)}\cr
&=& \int_\CX \;\exp\left(-2iJ+\,2i\, \sum_{k\geq 1}
\frac{(-1)^k(2k)!\zeta(2k+1)}{(2\pi)^{2k}}ch_{2k+1}(\CX)\right)\ .
\end{eqnarray}
Here, the ``even" part of the two Gamma classes cancel out
completely, suggesting that they, but not ``odd" parts, carry
RR-charge information. For Calabi-Yau 3-fold, the piece
$\int_\CX ch_3(\CX)$ is proportional to the Euler number and
represents exactly the quantum shift of the volume that
has been seen in the mirror map \cite{Jockers:2012dk,Halverson:2013qca}. This viewpoint also
conforms with the fact that there is no modification for Calabi-Yau
2-fold (times remaining flat directions), for which the ten-dimensional
spacetime theory has as many as 16 supercharges.

The ambiguity in determining RR-charge from the anomaly inflow
remains, as the D-brane and the I-brane inflow mechanisms always
conjugate one of the two factors as in
\eqref{OOB}.\footnote{Although, in principle, the Chern-Simons coupling
may be computable by direct string world-sheet method along the line of Refs.~\cite{Craps:1998fn,hep-th/9812071,hep-th/9812088,Craps:1998tw},
which had confirmed the first few terms of Refs.~\cite{GHM,CY}.}
However, once we accept \eqref{QK} as the quantum
version of the exponentiated K\"ahler class,
this ambiguity is lifted, and we come back to the same old Chern-Simons
coupling to spacetime RR-tensors for D-branes and
Orientifold planes \cite{Minasian:1997mm,GHM,CY,hep-th/9812071,hep-th/9707224,hep-th/9709219,Kim:2012wc}.

\vskip 2cm
\centerline{\bf Acknowledgement}
\vskip 1mm
We benefited from discussions and communications with
Dongmin Gang, Jaume Gomis, Jeff Harvey, Kentaro Hori, Emil Martinec,
Takuya Okuda, and Savdeep Sethi.
S.L. thanks the Perimeter Institute for hospitality where part of
this work has been done.
This work is supported in part (H.K.) by  NRF(National Research Foundation of Korea)
Grant funded by the Korean Government (NRF-2011-Global Ph.D. Fellowship Program),
and also (S.L.) by  the Ernest Rutherford fellowship of the Science and
Technology Facilities Council ST/J003549/1.

\vskip 2cm

\newpage
\appendix

\centerline{\bf\Large Appendix}
\section{Spherical Harmonics}\label{harmonics}
We summarize basic facts about the (monopole) spherical harmonics. In order
to discuss the projection condition under the parity, it is convenient to choose
a gauge where the monopole background vector field takes the following form
\begin{align}
  A = - \frac{B}{2}\cos\th d\varphi\ ,
\end{align}
valid in the region $0<\th<\pi$. In addition, we also need to choose a gauge
for the spin connection, as it affects the harmonics for spinors and vectors.
Our choice,
\begin{align}
  w^{\hat \theta}_{\;\; \hat \phi} = - \cos\th d\varphi\ ,
\end{align}
is such that spinor spherical harmonics are antiperiodic
along $\phi\rightarrow \phi+2\pi$.

The scalar monopole harmonics $Y_{\mathbf{q},jm}$ with $\mathbf{q}=\frac{B}{2}Q$ satisfy
\begin{align}
  -D_m^2 Y_{\mathbf{q},jm} = j ( j + 1) - \mathbf{q}^2 \ ,
  \qquad j = l + |\mathbf{q}| \ (l=0,1,2,..)\ ,
\end{align}
where the covariant derivative denotes
\begin{align}
  D = d - i Q A\ .
\end{align}
For later convenience, we present an explicit expression of the scalar monopole harmonics below,
\begin{align}
  Y_{\mathbf{q},jm}(\th,\varphi) =  M_{\mathbf{q},jm} (1-x)^{\a/2} (1+x)^{\b/2}
  P_n^{\a\b}(x) e^{im\varphi}\ ,
\end{align}
with
\begin{align}
  x = \cos\th\ , \ \a = - \mathbf{q} - m \ , \ \b=  \mathbf{q} - m\ , \
  n = j + m \ ,
\end{align}
and
\begin{align}
  M_{\mathbf{q},jm} = 2^m \sqrt{\frac{2j+1}{4\pi} \frac{(j-m)! (j+m)!}
  {(j-\mathbf{q})! (j+\mathbf{q})!}}\ .
\end{align}
Here the Jacobi polynomial $P_{n}^{\a\b}(x)$ is defined by
\begin{align}
  P_n^{\a\b}(x) = \frac{(-1)^n}{2^n n!} (1-x)^{-\a} (1+x)^{-\b} \frac{d^n}{dx^n}
  \Big[ (1-x)^{\a+n} (1+x)^{\b+n} \Big] \ .
\end{align}
Using the fact that
\begin{align}
  P_n^{\a\b}(-x) = (-1)^n P_n^{\b\a}(x) \ ,
\end{align}
it is straightforward to show that, for $0<\th<\pi$,
\begin{align}
  Y_{\mathbf{q}, jm}(\pi-\th,\pi+\varphi) = & (-1)^n e^{i\pi m} \, Y_{-\mathbf{q},jm}(\th,\varphi)
  \nonumber \\ = & (-1)^{l} e^{-i\pi|\mathbf{q}|} \, Y_{-\mathbf{q},jm}(\th,\varphi)\ .
\end{align}
For instance,
\begin{align}
  Y_{\pm \frac12, jm} (\pi-\th,\pi+\varphi) = (-i) (-1)^{l} Y_{\mp \frac12,jm}(\th,\varphi)\
  \text{for } j = l +\frac12 \ .
  \label{prop1}
\end{align}
The complex conjugate of the monopole harmonics satisfy
the following two relations,
\begin{align}
  Y_{\mathbf{q},jm}^*(\th,\varphi) = (-1)^{\mathbf{q}+m} Y_{-\mathbf{q}, j(-m)}(\th,\varphi)\ ,
\end{align}
and
\begin{align}
  \int_{S^2} \ Y_{\mathbf{q},jm}^*(\th,\varphi) Y_{\mathbf{q'},j'm'}(\th,\varphi)
  = \d_{qq'}\d_{jj'}\d_{mm'}\ .
\end{align}

We now move on to the spinor monopole harmonics.
It is useful to consider the eigenmodes $\Psi^\pm_{\mathbf{q},jm}$
of a modified Dirac operator
\begin{align}
  - i \g^3 \g^m D_m \Psi^\pm_{\mathbf{q},jm} = i \l_\pm  \Psi^\pm_{\mathbf{q},jm}
  \ , \qquad
  \l_\pm = \pm \sqrt{\left( j + \frac12 \right)^2 - \mathbf{q}^2 }\ ,
\end{align}
where
\begin{align}
  \Psi^\pm_{\mathbf{q},jm} = \begin{pmatrix}   Y_{\mathbf{q}-\frac 12,jm}
  \\ \pm Y_{\mathbf{q} + \frac 12, jm}  \end{pmatrix}\ .
\end{align}
Here the covariant derivative is
\begin{align}
  D = d - i Q A + \frac14 \w_{ab} \g^{ab}\ .
\end{align}
Using the property of the monopole harmonics (\ref{prop1}), one can show
\begin{align}
  \Psi_{\mathbf{q}=0,jm}^\pm ( \pi-\th, \pi +\varphi) = \mp i (-1)^{l}
  \begin{pmatrix} \pm Y_{\frac 12, jm}(\th,\varphi) \\
  Y_{ - \frac 12, jm}(\th, \varphi) \end{pmatrix}\ ,
  \label{prop2}
\end{align}
with $0<\th<\pi$.

Finally let us discuss about the one-form spherical harmonics defined by
\begin{align}
  C^1_{jm} = & + \frac{1}{\sqrt{j(j+1)}} d Y_{lm}\ ,
  \nonumber \\
  C^2_{jm} = & - \frac{1}{\sqrt{j(j+1)}}
  \ast d Y_{jm}\ ,
\end{align}
where $j\geq1$. Useful properties of the vector spherical harmonics can be
summarized as follow,
\begin{align}
  \ast C^2_{jm} = C^1_{jm}\ , \qquad
  \ast d C^2_{jm} =  \sqrt{j(j+1)} Y_{lm}\ , \qquad
  \ast d C^1_{jm} = 0 \ ,
\end{align}
which lead to
\begin{align}
  \ast d \ast d C^2_{jm} = & - j(j+1) C^2_{jm}\ ,
  \nonumber \\
  \ast d \ast d C^1_{jm} = & \ 0 \ .
\end{align}
Under the parity action, they transform as
\begin{align}
  C^1_{jm}(\pi-\th,\pi+\varphi) = & \ (-1)^j C^1_{jm}(\th,\varphi) \ ,
  \nonumber \\
  C^2_{jm}(\pi-\th,\pi+\varphi)  = & \ (-1)^{j+1} C^2_{jm}(\th,\varphi) \ .
\end{align}
%
%
%

\section{One-Loop Determinant on $\mathbb{RP}^2_b$}\label{squashing}

We will show that the partition function on the squashed real projective
space $\mathbb{RP}^2_b$ is independent of the squashing parameter $b$.
This section largely relies on the discussion in \cite{Gomis:2012wy}.
For details, please refer to Appendix A of the reference.

To compute the one-loop determinant around the SUSY saddle points, it is not
necessary to know all the eigenmodes of boson and fermion kinetic operators.
This is because, as we see in section 3, the huge cancelation between
boson and fermion eigenmodes occurs. It is therefore sufficient
to understand how the boson and fermion eigenmodes are paired by the supersymmetry.

\subsection{Chiral multiplet}

We start with a chiral multiplet of unit $U(1)$ gauge charge. To simplify the computation,
we choose a $\CQ$-exact regulator
\begin{align}
  \CL_\text{reg}= -\d_\e \d_{\bar \e} \Big[ \bar\psi \g^3 \psi - 2 \bar\phi \s_2 \phi \Big]\ ,
\end{align}
different to the one used in the main context. The above choice leads to the kinetic operators
around the saddle points (\ref{saddle1}), (\ref{saddle2})
\begin{align}
  \D_b = & - D_m^2 + \s^2 + \frac{q}{4} \CR + \frac{q-1}{f} v^mD_m +
  + \frac{q^2-2q}{4f^2}\ ,
  \nonumber \\
  \D_f = & - i \g^mD_m - \s\g^3 - i\frac{1}{2f} \g^3 + i \frac{q-1}{2f} v_m \g^m
  + i \frac{q-1}{2f} w\ ,
\end{align}
where the covariant derivative involves the background gauge field $V$ given in (\ref{Vr}),
\begin{align}
  D_m \phi = & \left( \partial_m - i A_m + i q V_m \right) \phi \ ,
  \nonumber \\
  D_m \psi = & \left( \partial_m - iA_m + \frac{1}{4}w_{ab} \g^{ab}
  + i (q-1) V_m \right) \psi\ ,
\end{align}
and
\begin{align}
  v^m = \bar \e \g^m \e \ , \qquad w = \bar \e \e\ .
\end{align}
Here $\CR$ denotes the scalar curvature of $\mathbb{RP}^2$.
As in section 3, it is convenient to consider spinor eigenmodes for
an operator $\g^3\D_f$ instead of $\D_f$.

One can show that there is a pair between a scalar eigenmode for
$\D_b\doteq - M(M+2\s)$ and two spinor eigenmodes for $\g^3\D_f\doteq M,
-(M+2\s)$, subject to either (\ref{chiralproj}) or
(\ref{twistedchiralproj}) projection conditions.
The precise map which pairs the scalar and spinor
eigenmodes is the following; Given a spinor eigenmode $\Psi$ for $\g^3\D_f \doteq M$,
one can show that
\begin{align}
  \bar \e \Psi
  \label{map1}
\end{align}
is a scalar eigenmode for $\D_b \doteq -M(M+2\s)$.
On the other hand, one can define a pair of spinors
\begin{align}
  \Psi_1 = \g^3\e\Phi\ , \qquad \Psi_2 = i\g^m \e D_m \Phi + \g^3 \e \left(\s \Phi +
  i \frac{q}{2f} \right) \Phi\ ,
\end{align}
where $\Phi$ is a scalar eigenmode  for $\D_b \doteq - M ( M + 2\s )$. One can show that
\begin{align}
  M \Psi_1 + \Psi_2 \ , \qquad - (M+2\s) \Psi_1 + \Psi_2
  \label{map2}
\end{align}
are the eigenmodes for $\g^3\D_f \doteq M$ and $\g^3\D_f \doteq -(M+2\s)$ respectively.

Any modes in such a pair can not contribute to the one-loop determinant due to the
cancelation. As a consequence, the nontrivial contributions arise from the eigenmodes
where either the map (\ref{map1}) or the map (\ref{map2}) becomes ill-defined.

\paragraph{Unpaired spinor eigenmode} If a spinor eigenmode vanishes when
contracted with $\bar \e$, there is no scalar partner. Such an unpaired spinor
eigenmode takes the following form
\begin{align}
  \Psi = e^{- i J\varphi} h(\th) \bar \e  \ ,
\end{align}
where
\begin{align}
  i J = \left( M l + \s l + i \frac{q-2}{2} \right) \ ,
\end{align}
and
\begin{align}
  \frac 1f \partial_\th h = \tan\th \left( \frac Jl - \frac{q-2}{2l} + i \frac{q-2}{2f}
  \right) h \ .
\end{align}
For the normalizability, one has to require $J$ to be non-negative.
Note that the function $h(\th)$ is even under the parity, i.e., $h(\th)=h(\pi-\th)$.
One can show that
$J$ should be further restricted to be even (odd) to satisfy the projection conditions
in the even (odd) holonomy, i.e.,
\begin{align}
  M l = &  i \left( 2k + 1+ i \s l - \frac q2 \right) \text{ for even holonomy}\ ,
  \nonumber \\
  M l = & i \left( 2k + 2 + i \s l- \frac q2 \right) \text{ for odd holonomy}\ ,
  \label{result1}
\end{align}
with $k \geq 0$.

\paragraph{Missing spinor eigenmode}

Suppose that a scalar eigenmode $\Phi$ for $\D_b \doteq - M (M+2\s)$
fails to provide two independent spinor eigenmodes via the map (\ref{map2}).
It happens when
\begin{align}
  \Psi_2 = - M \Psi_1\ ,
\end{align}
which leads to a missing spinor eigenmode for $\g^3\D_f \doteq M$.
One can verify that such a scalar eigenmode $\Phi$ missing a spinor eigenmode takes the
following form
\begin{align}
  \Phi = e^{i J \varphi} \chi(\th) \ ,
\end{align}
where
\begin{align}
  iJ = - \left( M l + \s l + i \frac q2 \right)\ ,
\end{align}
with $J \geq 0$ for the normalizability, and
\begin{align}
  \frac 1f \partial_\th \chi = \tan \th \left( \frac Jl + \frac{q}{2l} -
  \frac{q}{2f} \right) \chi \ .
\end{align}
To satisfy the projection condition in the even (odd) holonomy,
one can show that
\begin{align}
  M l = & - i \left( 2k - i \s l + \frac q2\right) \text{ for even holonomy }\ ,
  \nonumber \\
  M l = & - i \left( 2k+1 - i \s l + \frac q2\right) \text{ for odd holonomy }\ ,
  \label{result2}
\end{align}
with $k \geq 0$.

\paragraph{One-loop determinant} Combining all the results (\ref{result1}) and
(\ref{result2}), one can show
\begin{align}
  \frac{\det \D_f}{\det \D_b} \simeq & \frac{\det \g^3\D_f}{\det \D_b}
  \simeq  \left\{ \begin{array}{cc}
  \prod_{k\geq 0} \frac{2k + 1+ i \s l - \frac q2 }{ 2k - i \s l + \frac q2}
  & \text{for even holonomy}\\
  \prod_{k\geq 0} \frac{2k + 2+ i \s l - \frac q2 }{ 2k +1 - i \s l + \frac q2}
  & \text{for odd holonomy}\end{array} \right. \ ,
\end{align}
where the symbol $\simeq$ represents the equality up to a sign independent of $\s$.
From the comparison to the results in section 3, one can fix the sign factor by the unity.
These results are in perfect agreement to those for $\mathbb{RP}^2$.

\subsection{Vector multiplet}
We now in turn compute the one-loop determinant from the vector multiplet.
Denoting the various fluctuation fields as follows
\begin{align}
  A = A_\text{flat} + a \ , \qquad \s_1 = \zeta\ , \qquad \s_2 = \s + \eta\ ,
\end{align}
let us decompose all the adjoint fields $(a,\z,\eta)$ into Cartan-Weyl basis. From now on,
we focus on the W-boson of charge $\a$, a root of $G$, and its super partners.
The kinetic Lagrangian for the vector multiplet is chosen as a $\CQ$-exact
regulator.

As explained in \cite{Benini:2012ui} and \cite{Gomis:2012wy} that the four bosonic modes contain two
longitudinal modes with $a\sim D \eta$ that correspond to a gauge rotation and the volume of
the gauge group $G$. Using the standard Fadeev-Popov method, one can argue that these longitudinal
modes can not contribute to the one-loop determinant. Thus we need to find how two transverse
modes with $\ast D \ast a = 0$ can be paired with spinor eigenmodes.

The kinetic operators of our interest are
\begin{align}
  \D_b = & \begin{pmatrix} - \ast d \ast d + (\a\cdot \s)^2 & - \ast d \frac{1}{f} \\
  + \frac1f \ast d & - \ast d \ast d + \frac{1}{f^2} + (\a\cdot \s)^2 \end{pmatrix}\ ,
  \nonumber \\
  \D_f = & i \g^m D_m + (\a\cdot \sigma)\g^3\ ,
\end{align}
with the gauge choice $\ast d\ast a = 0$. The operator $\D_b$ acts on the fluctuation
fields $(a,\zeta)$ subject to the projection conditions (\ref{vectorproj}) for the even holonomy and
the twisted projection conditions for the odd holonomy.
Instead of $\D_b$, it is convenient to consider the following operator
\begin{align}
  \d_b \equiv \begin{pmatrix} i \a\cdot \s & - \ast d \\ \ast d & \frac 1f + i \a\cdot \s
  \end{pmatrix}\ .
\end{align}
One can show that the operator $\d_b$ satisfies the relation $\d_b^2 = \D_b + 2i (\a\cdot \s)\d_b$,
or equivalently,
\begin{align}
  \d_b \doteq - i M\ , + i (M+2\a\cdot \s) \ \leftrightarrow \ \D_b \doteq - M ( M + 2\a\cdot \s )\ .
\end{align}

Let $(\CA,\S)$ and $\L$ be bosonic eigenmodes for $\d_b \doteq -i M $ and
fermionic eigenmodes for $\g^3\D_f \doteq - M$. They can be shown to be mapped to
each other by
\begin{align}
  \CA = - i \left( M + \a\cdot\s\right) \bar \e \g_m \L e^m - d \left( \bar \e\g^3\L \right) \ ,
  \qquad
  \S = (M + \a \cdot \s ) \bar \e \L \ ,
  \label{map3}
\end{align}
and
\begin{align}
  \L = \left( \g^3\g^m \CA_m + i \S \g^3 \right) \e \ .
  \label{map4}
\end{align}
Again, one can have nontrivial contribution to the one-loop determinant
from either unpaired or missing spinor eigenmodes.

\paragraph{Unpaired spinor eigenmodes} An unpaired spinor eigenmode,
annihilated by the map (\ref{map3}), takes the following form
\begin{align}
  \L = e^{-i J\varphi} h(\th) \bar \e \ ,
\end{align}
where
\begin{align}
  i \left( J + 1\right) = M l + \a \cdot \s l\ ,
\end{align}
with $J\geq 0$ due to the normalizability, and
\begin{align}
  \frac 1f \partial_\th h + \tan\th \left( \frac1f - \frac{J+1}{l} \right) h = 0 \ .
\end{align}
Note that the function $h(\th)$ is even under the parity, $h(\pi-\th) = h(\th)$.
In order to satisfy the projection conditions in the even (odd) holonomy,
the non-negative integer $J$ should be further constrained to be odd (even), i.e.,
\begin{align}
  M l = & i \left( 2k+2 + i\a\cdot\s \right) \text{ for even holonomy}\ ,
  \nonumber\\
  M l = & i \left( 2k + 1 + i \a\cdot\s \right) \text{ for odd holonomy}\ ,
  \label{result3}
\end{align}
with $k\geq 0$.

\paragraph{Missing spinor eigenmodes} One can show from the map (\ref{map4}) that
a bosonic eigenmode with missing spinor partner can take the following form
\begin{align}
  \CA = e^{i(J+1)\varphi} \chi(\th) \left(e^1 + i \cos\th \e^2 \right) \ ,
  \qquad
  \S = i e^{i(J+1)\varphi} \chi(\th) \sin\th\ ,
\end{align}
where $e^m$ denotes the vielbein of $\mathbb{RP}^2_b$, and
\begin{align}
  i (J + 1) = - \left( M l + \a\cdot\s l \right)\ ,
\end{align}
and
\begin{align}
  \frac 1f \partial_\th \chi + \tan\th \left( \frac{1}{f} - \frac{J+1}{l} \right) \chi = 0 \ .
\end{align}
The normalizability requires $J$ to be non-negative. The projection conditions in even (odd)
holonomy are satisfied if $J$ are even (odd), i.e.,
\begin{align}
  M l = & -i \left( 2k + 1 - i\a\cdot\s \right) \text{ for even holonomy}\ ,
  \nonumber\\
  M l = & - i \left( 2k + 2 - i \a\cdot\s \right) \text{ for odd holonomy}\ ,
  \label{result4}
\end{align}
with $k\geq 0$.

\paragraph{One-loop determinant} Collecting all the results (\ref{result3}) and
(\ref{result4}), the one-loop determinant from the vector multiplet becomes
\begin{align}
  \frac{\det \D_f}{\sqrt{\det \D_b}} \simeq
  \frac{\det \g^3\D_f}{\det \d_b} \simeq & \left\{ \begin{array}{cc}
  \prod_{\a \in \D}\prod_{k\geq 0} \frac{2k+2 + i\a\cdot\s}{2k+1 - i \a\cdot\s}
  & \text{for even holonomy} \\
  \prod_{\a \in \D}\prod_{k\geq 0} \frac{2k+1 + i\a\cdot\s}{2k+2 - i \a\cdot\s}
  & \text{for odd holonomy}
  \end{array}\ . \right.
\end{align}
By comparing the results to those in section 3, one can fix the sign factor by the unity.
Again, these results perfectly agree with those for $\mathbb{RP}^2$.

\section{Spin$^c$ Structure and Tachyon Condensation}

The central charge of the D-branes is recently revisited from the exact hemisphere
partition function computation \cite{Hori:2013ika,Honda:2013uca}, which shows
that $\hat \Gamma_c$ class replaces
the $\CA^{1/2}$ class. In this note, we computed central charges of Orientifold planes
and show how $L^{1/2}$ class in the traditional central charge formula must be
similarly modified in terms of $\CA$  class and $\hat \Gamma_c$ class. The original
formulae for the RR-charges and the central charges were motivated by anomaly inflow,
which can potentially miss some of $(4k+2)$-form pieces in the exponent.

An odd fact is that the partition function computations, with Dirichlet condition explicitly
imposed, also seem to miss a factor $e^{-c_1(\CN)/2}$ that is necessary when $\CM$ is a proper
submanifold of $\CX$ and is not Spin but only Spin$^c$.
Here, we will outline how this factor re-emerges, at least, from the viewpoint
of tachyon condensation. After presenting  results in Refs.~\cite{Hori:2013ika,Honda:2013uca},
we compute the central charge of a lower-dimensional D-brane wrapping a hypersurface
in the Calabi-Yau space, with careful consideration given to charge assignment of
the vacua. From this computation, the subtle factor $e^{ -c_1(\CN)/2}$ is rescued.

In order to preserve the supersymmetry chosen as in (\ref{killing}) on the northern hemisphere,
one has to add the Chan-Paton factor
\begin{align}
  \text{Tr}_\CV \Big[ \CP \exp\left( - i \int d\varphi \ \CA_\varphi \right) \Big]
\end{align}
with
\begin{align}
  \CA_\varphi = & \rho_\ast\left( A_\varphi + i \s_2 \right)  - \frac{r_\ast}{2r}
  - i \Big\{ \CQ, \bar \CQ \Big\}
 \nonumber \\ &
 + \frac{1}{\sqrt2}  \left( \psi_+^i -\psi_-^i \right) \partial_i \CQ
  +  \frac{1}{\sqrt2} \left( \bar \psi_+^i - \bar \psi_-^i \right) \partial_i \bar \CQ \ .
  \label{CP1}
\end{align}
Here $\CV$ denotes a $Z_2$ graded Chan-Paton vector space.
The tachyon profile $\CQ(\phi)$ is an operator acting on the
vector space $\CV$, anti-commuting with fermions, and obeys
the following relation
\begin{align}
  Q^2 = \CW \cdot {\bf 1}_\CV\ ,
\end{align}
where $\CW(\phi)$ denotes a given superpotential. The $G \times U(1)_v$
representation of the Chan-Paton vector space $\CV$ is specified by $\rho_\ast$ and $r_\ast$,
\begin{align}
  \rho(g) Q(\phi) \rho(g)^{-1} = & \ Q ( g\phi)\ ,
  \nonumber \\
  \l \cdot \l^{r_\ast} Q(\phi) \l^{-r_\ast} = & \ Q ( \l^q \phi)\ ,
\end{align}
where $g\in G$.
The exact hemisphere partition function can be expressed by
\begin{align}
  Z_{D^2} \propto \frac{1}{\left| W(G) \right|} \int_{\mathfrak{t}} d \s
  \ e^{-2\pi i \xi_\text{ren} \text{tr} \s - \th \text{tr} \s } \times \text{Tr}_\CV\Big[ e^{2\pi \r_\ast(\s) + i \pi r_\ast} \Big]
  \times Z_\text{1-loop}
  \label{centralchargef3}
\end{align}
with
\begin{align}
  Z_\text{1-loop} =
  \prod_{\a>0} \Big[\a\cdot \s \sinh{\a\cdot\s}\Big]
  \prod_{w_a \in\mathbf{R^a}} \G\left( \frac{q_a}{2} - i w_a \cdot \s \right)\ ,
\end{align}
where $\mathfrak{t}$ is the Cartan subalgebra of the gauge group $G$,
and $W(G)$ is the Weyl group.

In what follows, we consider the $U(1)$ GLSM describing the degree $N$
hypersurface of $\mathbb{CP}^{N-1}$ studied in section 5 for simplicity.
\begin{align}
  \CW = P G_N(X^i) \ ,
\end{align}
where $G_N(X^i)$ denotes a homogeneous polynomial of degree $N$. This model describes
the non-linear sigma model whose target space is a CY hypersurface $X$ in $\mathbb{CP}^{N-1}$.

Taking into account for the Kn\"orrer map to relate the GLSM brane $\mathfrak{B}_{UV}$
to the NLSM brane $\mathfrak{B}_{IR}$ \cite{Herbst:2008jq},
one can show that
\begin{align}
  ch\big[\mathfrak{B}_{IR}\big] =
  \frac{1}{1 - e^{- 2\pi i N ( q/2- i \s) } } \times \text{Tr}_{\CV}
  \Big[ e^{2\pi \r_\ast(\s) + i \pi r_\ast} \Big]\ ,
\end{align}
where $\CV$ denotes the Chan-Paton vector space of $\mathfrak{B}_{UV}$.
Note that the Kn\"orrer map also leads to the shift of the theta angle
\begin{align}
  \theta_{UV} = \theta_{IR} - \pi N\ .
\end{align}
The central charge of the NLSM brane $\mathfrak{B}_{IR}$ then takes the following form
\begin{align}
  Z\left(\mathfrak{B}_{IR}\right) =
  2i\pi \beta \int_{q/2-i\infty}^{q/2+i\infty} \frac{d\e}{2\pi i} \
  e^{2\pi \xi \e - i \theta_{IR} \e} \times \frac{N}{\e^{N-1}} \times \frac{ \G( 1+ \e)^N}{\G(1+ N\e)}  \times
  ch\big[\mathfrak{B}_{IR}\big]
  \label{centralchargef4}
\end{align}
with $\beta=(r\Lambda)^{c/6}/(2\pi)^{(N-2)/2}$.
Focusing on the perturbative part of the central charge,
one can finally obtain the large-volume expression \cite{Hori:2013ika,Honda:2013uca}
\begin{align}
 Z_\text{pert}\left(\mathfrak{B}_{IR}\right)  = &
  \left(2\pi i\right)^{N-2}\beta  \int_X \ e^{-i\xi H - \frac{\th_{IR}}{2\pi} H} \times
 \frac{\G(1+ \frac{H}{2\pi i})^N}{\G(1+\frac{N H}{2\pi i})} \times ch\big[ \mathfrak{B}_{IR} \big]
 \nonumber \\ = & \left(2\pi i\right)^{N-2}\beta   \int_X \ e^{-iJ- B} \wedge \hat \G_c(X)
 \wedge ch\big[ \mathfrak{B}_{IR} \big] \ ,
 \label{largevolcharge}
\end{align}
where $H$ is the hyperplane class of $\mathbb{CP}^{N-1}$.

To be more concrete let us consider a tachyon profile $\CQ$
\begin{align}
  \CQ = X^a \eta_a + P \tilde \eta + G_N \bar{\tilde \eta}\ ,
\end{align}
where the fermionic oscillators satisfy the following anti-commutation relations
\begin{align}
  \{\tilde \eta, \bar{\tilde \eta} \} = 1\ , \qquad
  \{\eta_a, \bar \eta_b\}=\d_{ab}
\end{align}
with $a=1,2,..,n$.
Since the boundary potential becomes
\begin{align}
  \Big\{ \CQ, \bar \CQ \Big\} = \sum_{a=1}^n |X^a|^2 + |P|^2 + |G_N(X^i)|^2\ ,
\end{align}
the above tachyon profile describes a lower-dimensional brane
wrapping a submanifold at $X^a=0$ in the Calabi-Yau space $X$ in the geometric phase.
One can easily show that the Chern character of the brane $\mathfrak{B}_{IR}$
is
\begin{align}
  ch\big[\mathfrak{B}_{IR}\big]= e^{-\pi i n\e} \Big(2i \sin( \pi \e)\Big)^n\ ,
\end{align}
where $\e=q/2-i\s$. Then, the central charge of the brane
in the large volume limit (\ref{largevolcharge})  can be written as
\begin{align}
  Z_\text{pert}\left(\mathfrak{B}_{IR}\right) =
& \left(2i \pi \right)^{N-1} \b \int_X
  e^{-i\xi H - \frac{\th_{IR}}{2\pi} H} \times
  \frac{\G( 1+ \frac{H}{2 \pi i})^{N-n}}{\G(1-\frac{H}{2\pi i})^n \G(1+ \frac{N H}{2\pi i})}
  \times H^n \times e^{ - \frac{nH}{2} }
  \nonumber \\ = &
  \left(2i \pi \right)^{N-1}   \b \int_X
  e^{-i J - B} \wedge
  \frac{\hat \G_c (\CT)}{\hat \G_c(-\CN)} \wedge e(\CN) \wedge e^{ - \frac12 c_1(\CN)}\ .
  \label{cccc}
\end{align}
Note that one can see the very subtle factor $e^{-c_1(\CN)/2}$ emerges from the
partition function computation again. As a byproduct, we also confirmed that the overall
normalization factor $(2\pi i)^{N-1}\beta= (2\pi)^{N/2}i^{N-1}(r\Lambda)^{c/6}$
are the same for any dimensional D-branes,
which is consistent with the tadpole comparison of the section 6 above.

What remains unclear
to us is  how this factor should be recovered when the Dirichlet boundary condition
is imposed from the outset.\footnote{A related issue was addressed in Ref.~\cite{Honda:2013uca}.}
For D-branes here,  the result (\ref{cccc})
obtained from the tachyon condensation exactly agrees with the central charge
for a D-brane supporting an extra ``line bundle'' $\CO(-n/2)$,
which can be described by imposing the Dirichlet boundary condition on $X^a$
and adding Wilson loop of charge $\rho_\ast = - n/2$.
The line bundle $\CO(-n/2)$
accounts for the factor $e^{-\frac12 c_1(\CN)}$ canceling the Freed-Witten global anomaly.
As there is analog of neither tachyon condensation representation nor Wilson loop
for Orientifold planes, however, a better understanding of this issue is needed for addressing
central charges of all RR-charged objects.

\end{document}